\documentclass[a4paper,11pt]{article}
\usepackage{amsmath,amssymb,color,comment}
\usepackage{slashed, tensor, bm, physics}
\usepackage{caption}
\usepackage{graphicx}
\usepackage{multirow}
\usepackage{float}
\usepackage[compat=1.1.0]{tikz-feynhand}
\usepackage{jheppub} 

\usepackage[T1]{fontenc} 
\usepackage{booktabs} 
\newcommand{\drawsquare}[2]{\hbox{%
\rule{#2pt}{#1pt}\hskip-#2pt
\rule{#1pt}{#2pt}\hskip-#1pt
\rule[#1pt]{#1pt}{#2pt}}\rule[#1pt]{#2pt}{#2pt}\hskip-#2pt
\rule{#2pt}{#1pt}}
\newcommand{\Yfund}{\raisebox{-.5pt}{\drawsquare{6.5}{0.4}}}
\newcommand{\Yantifund}{\overline\Yfund}
\newcommand{\Ysymm}{\raisebox{-.5pt}{\drawsquare{6.5}{0.4}}\hskip-0.4pt%
        \raisebox{-.5pt}{\drawsquare{6.5}{0.4}}}
\newcommand{\Yasymm}{\raisebox{-3.5pt}{\drawsquare{6.5}{0.4}}\hskip-6.9pt%
        \raisebox{3pt}{\drawsquare{6.5}{0.4}}}
\newcommand{\Yantisymm}{\overline\Ysymm}
\newcommand{\Yantiasymm}{\overline\Yasymm}
\newcommand{\U}{{\rm U}}
\newcommand{\SU}{{\rm SU}}
\newcommand{\SO}{{\rm SO}}
\newcommand{\Sp}{{\rm Sp}}

\def\rsato#1{\textcolor{magenta}{[RS: #1]}}
\def\ST#1{\textcolor{orange}{[ST: #1]}}

\title{QCD axion from chiral gauge theories}
\author[a]{Ryosuke Sato,}
\author[b]{and Shonosuke Takeshita}

\affiliation[a\,]{Department of Physics, The University of Osaka, Toyonaka, Osaka 560-0043, Japan}
\affiliation[b\,]{Physics Program, Graduate School of Advanced Science and Engineering, Hiroshima University, Higashi-Hiroshima 739-8526, Japan}

\emailAdd{rsato@het.phys.sci.osaka-u.ac.jp}
\emailAdd{shonosuke@hiroshima-u.ac.jp}

\abstract{
  We present models of axion based on supersymmetric chiral gauge theories. In these models, the PQ symmetry is spontaneously broken by the non-perturbative dynamics of chiral gauge theory. Thanks to supersymmetry, IR dynamics of the models are calculable.
  We also present an example of a QCD axion model that is compatible with $\SU(5)$ grand unification. We find that in order to realize the gauge coupling unification with a certain precision, the GUT scale is the same with the PQ breaking scale, and the SUSY breaking scale is ${\cal O} (10^9)~{\rm GeV} $.
}

\begin{document} 
\begin{flushright}
OU-HET-1255\\
HUPD-2506
\end{flushright}
\maketitle
\flushbottom

\section{Introduction}
The strong CP problem \cite{Jackiw:1976pf, Callan:1976je} is one of the most mysterious puzzles in the Standard Model (SM). The effective $\theta$-angle in the QCD sector is severely constrained as $|\bar\theta|\lesssim 10^{-10}$ from the null observation of the neutron electric dipole moment (EDM) \cite{Abel:2020pzs}, though ${\cal O}(1)$ CP phase has been observed in the Cabibbo-Kobayashi-Maskawa (CKM) matrix. This would indicate some unknown mechanism to realize the hierarchy $|\delta_{\rm CKM}| \gg |\bar\theta|$ which is incorporated in physics beyond the standard model.
The QCD axion \cite{Peccei:1977hh, Peccei:1977ur, Weinberg:1977ma, Wilczek:1977pj} is one of the most interesting proposals to solve the strong CP problem. It is a pseudo Nambu-Goldstone (NG) boson associated with spontaneous breaking of global $\U(1)_{\rm PQ}$ symmetry which is anomalous under QCD interaction. Since the QCD axion couples to $\SU(3)_c$ gauge field via a coupling $(a/32\pi^2 f_a) G_{\mu\nu}\tilde G^{\mu\nu}$, the $\theta$-angle is ``promoted to a dynamical scalar field''. Then the $\theta$-angle is absorbed by the axion at its potential minimum \cite{Vafa:1984xg} and the predicted size of the neutron EDM becomes much smaller than the current experimental upperbound \cite{Georgi:1986kr}.

An interesting scenario for the QCD axion would be spontaneous $\U(1)_{\rm PQ}$ breaking induced by strong dynamics of a new gauge interaction~\cite{Kim:1984pt, Choi:1985cb}.
In this case, the order parameter of PQ symmetry breaking is a composite operator.
Refs.~\cite{Randall:1992ut,Dobrescu:1996jp,Redi:2016esr,Lillard:2017cwx,Lillard:2018fdt,Lee:2018yak,Gavela:2018paw,Vecchi:2021shj,Contino:2021ayn,Cox:2023dou, Nakagawa:2024kcb, Gherghetta:2025fip} utilize this property to solve the axion quality problem. 
Most models utilize strong dynamics of QCD-like gauge theories. On the other hand, discussion of the QCD axion based on strong dynamics of chiral gauge theories is quite limited \cite{Gavela:2018paw, Cox:2023dou} because of the difficulty of the analysis, as chiral gauge theories generally lack analytical tools for studying their non-perturbative dynamics.
Recently, refs.~\cite{Csaki:2021xhi, Csaki:2021aqv} have analyzed the strong dynamics of chiral gauge theories in supersymmetric (SUSY) models with small SUSY breaking \cite{Aharony:1995zh, Alvarez-Gaume:1996vlf, Alvarez-Gaume:1997wnu, Martin:1998yr, Cheng:1998xg, Strassler:1997ny, Kitano:2011zk, Murayama:2021xfj}. Thanks to SUSY, non-perturbative dynamics of chiral gauge theory can be solved in analytic expressions.

In this paper, we apply the analysis in \cite{Csaki:2021xhi, Csaki:2021aqv} to chiral gauge theories, and show some explicit examples of the QCD axion from chiral gauge theories.
This paper is organized as follows. 
In section \ref{sec:models}, we show explicit examples of models.
In section \ref{sec:GUT}, we show an example of model of the QCD axion which is compatible with $\SU(5)$ grand unification, and discuss its phenomenology and cosmology.
Section \ref{sec:summary} is devoted to conclusions and discussions.
In appendix \ref{sec:Wdyn}, we discuss a dynamical superpotential in $\SU(2N)$ chiral gauge theory.
In appendix \ref{sec:125GeV Higgs}, we show the calculation of the Higgs boson mass.
In appendix \ref{sec:RGEs}, we list the input parameters and RGEs for coupling constants.
In appendix \ref{sec:RGE proton decay}, we show the RGEs for proton decay operators.

\section{Axion from SUSY chiral gauge theories} \label{sec:models}
In this section, we present two simple toy models of chiral gauge theories that include an axion. This axion couples to the $\SU(N)$ gauge field through the following interaction:
\begin{align}
  {\cal L} = \frac{a}{f_a} \frac{1}{32\pi^2} G_{\mu\nu} \tilde G^{\mu\nu}.
\end{align}
Similar to the analysis presented in refs.~\cite{Csaki:2021xhi,Csaki:2021aqv}, where non-perturbative dynamics of chiral gauge theories were solved using supersymmetry, we utilize SUSY chiral gauge theories with small soft SUSY breaking.

\subsection{Axion from SUSY Georgi-Glashow type model}\label{sec:GG model}
\begin{table}
  \centering
\begin{tabular}{|c||c|c||c|c|c||c|}
  \hline
           & $\SU(2N+4)_1$      & $\SU(N)_2$  & $\U(1)_{\rm PQ}$ & $\U(1)_q$ & $\U(1)_{R}$ & $\SU(2N)_{\bar F}$\\\hline\hline
  $\bar F_q$ & $\Yantifund$ & $\Yfund$     &  $N+1$ & $1$ & $0$ & \multirow{2}{*}{\Yfund} \\\cline{1-6}
  $\bar F_{\bar q}$ & $\Yantifund$ & $\Yantifund$ & $N+1$ & $-1$ & $0$ & \\\hline
  $A$      & $\Yasymm$    & $\mathbf{1}$ & $-N$ & $0$ & $-3/(N+1)$ & $\mathbf{1}$ \\\hline
  \end{tabular}
  \caption{Matter content of SUSY Georgi-Glashow model.}\label{tab:axion from GG model}
\end{table}

We introduce $\SU(2N+4)_1 \times \SU(N)_2$ gauge symmetry and 
chiral multiplets
$\bar F_q ~(\Yantifund, \Yfund)$,
$\bar F_{\bar q} ~(\Yantifund, \Yantifund)$,
and $A ~(\Yasymm, \mathbf{1})$.
We assume that the dynamical scale of $\SU(N)_2$ gauge symmetry is much below than that of $\SU(2N+4)_1$ gauge symmetry, i.e., $\Lambda_N \ll \Lambda_{2N+4}$. We assume the tree level superpotential $W_{\rm tree}$ to be zero. In this setup, we can find $\U(1)_{\rm PQ} \times \U(1)_q \times \U(1)_{R}$ global symmetries, which are free from $\SU(2N+4)_1$ anomaly. The matter content and the charge assignment under the global symmetries are summarized in table~\ref{tab:axion from GG model}. 
Note that $\SU(2N)_{\bar F}$ global symmetry arises in the limit of $\Lambda_N\to 0$ and this $\SU(2N)_{\bar F}$ global symmetry is equivalent to that of \cite{Csaki:2021xhi}. $\SU(N)_2 \times \U(1)_q$ can be understood as a subgroup of $\SU(2N)_{\bar F}$.

Let us discuss the symmetry breaking pattern in the $D$-flat direction.
We write $A$ as $(2N+4) \times (2N+4)$ matrix and $\bar F_q$ and $\bar F_{\bar q}$ as $N \times (2N+4)$ matrix.
$\SU(2N+4)_1 \times \SU(N)_2$ gauge transformation for $A$, $\bar F_q$, and $\bar F_{\bar q}$ is given as
\begin{align}
  A \to U A U^T, \qquad
  \bar F_q \to U^* \bar F_q V^T, \qquad
  \bar F_{\bar q} \to U^* \bar F_{\bar q} V^\dagger, \label{eq:GG gauge transf}
\end{align}
where $U \in \SU(2N+4)_1$ and $V \in \SU(N)_2$.
$D$-flat condition is
\begin{align}
  2A^\dagger A - \bar F_q \bar F_q^\dagger - \bar F_{\bar q} \bar F_{\bar q}^\dagger \propto I_{2N}, \qquad
  \bar F_q^\dagger \bar F_q - \bar F_{\bar q}^T \bar F_{\bar q}^* \propto I_N.
\end{align}
As a solution, we obtain
\begin{align}
  A = \left(\begin{array}{cccc}
    0_{N\times N} & -ic I_N & & \\
    ic I_N & 0_{N\times N} & & \\
    && 0_{2\times 2} & -ic'I_2 \\
    && ic' I_2 & 0_{2\times 2}
  \end{array}\right), \quad
  \bar F_q = \phi \left(\begin{array}{c}
    I_{N} \\
    0_{N\times N} \\
    0_{N\times 4} \\
  \end{array}\right), \quad
  \bar F_{\bar q} = \phi \left(\begin{array}{c}
    0_{N\times N} \\
    I_{N} \\
    0_{N\times 4} \\
  \end{array}\right),
  \label{eq:vev GG model}
\end{align}
where $c$, $c'$, and $\phi$ are constants that satisfy $|c|^2 = |\phi|^2 + |c'|^2$.
Let us identify the unbroken gauge symmetry in this direction.
The VEVs given in eq.~(\ref{eq:vev GG model}) are invariant under a gauge transformation eq.~(\ref{eq:GG gauge transf}) with
\begin{align}
  U = \left(\begin{array}{ccc}
    \tilde V_{\SU(N)} &&\\
    & \tilde V_{\SU(N)}^* & \\
    && I_4
  \end{array}\right), \qquad
  V = \tilde V_{\SU(N)},
  \label{eq:SU(N) gauge transf}
\end{align}
and
\begin{align}
  U = \left(\begin{array}{cc}
    I_{2N} & \\
    & \tilde U_{\Sp(4)}
  \end{array}\right), \qquad
  V = I_N. \label{eq:Sp(4) gauge transf}
\end{align}
Note that $\tilde V_{\SU(N)} \in \SU(N)_d$ and $\tilde U_{\Sp(4)} \in \Sp(4)_1$.
$\SU(2N+4)_1$ gauge group contains $\SU(N)_1 \times \Sp(4)_1$ as a subgroup.
The gauge transformation eq.~(\ref{eq:SU(N) gauge transf}) shows $\SU(N)_d$ which is a diagonal subgroup of $\SU(N)_1 \times \SU(N)_2$ is unbroken. Also, the gauge transformation eq.~(\ref{eq:Sp(4) gauge transf}) shows $\Sp(4)_1 \in \SU(2N+4)_1$ is unbroken.
The $\SU(N)_d$ gauge coupling is determined by the following matching condition at tree level:
\begin{align}
  \frac{1}{\alpha_{\SU(N)_d}} = \frac{1}{\alpha_{\SU(2N+4)_1}} + \frac{1}{\alpha_{\SU(N)_2}}.
\end{align}

The VEVs eq.~(\ref{eq:vev GG model}) spontaneously break both $\U(1)_{\rm PQ}$ and $\U(1)_{R}$ symmetry. On the other hand, we can find the VEVs are invariant under the following combination of $\U(1)_q$ transformation,
\begin{align}
  \bar F_{q} \to e^{i\alpha} \bar F_q, \qquad
  \bar F_{\bar q} \to e^{-i\alpha} \bar F_{\bar q},
\end{align}
and $\U(1)_1 (\subset \SU(2N+4)_1) $ gauge transformation eq.~(\ref{eq:GG gauge transf}) with
\begin{align}
  U = \left(\begin{array}{ccc}
    e^{i\alpha} I_{N} && \\
    & e^{-i\alpha} I_{N} & \\
    && I_{4}
  \end{array}\right), \qquad
  V = I_{N}.
\end{align}
Thus, the VEVs eq.~(\ref{eq:vev GG model}) are invariant under the $\U(1)_d$ transformation which is a diagonal subgroup of $\U(1)_q \times \U(1)_1$.
To summarize, symmetry breaking pattern by the VEVs eq.~(\ref{eq:vev GG model}) is
\begin{align}
  [\SU(2N+4)_1 \times \SU(N)_2] \times \U(1)_{\rm PQ} \times \U(1)_q \times \U(1)_{R} ~\to~ [\Sp(4)_1 \times \SU(N)_d] \times \U(1)_d.
\end{align}
Here, the square brackets $[\cdots]$ indicate the gauge symmetries, while the terms outside the brackets represent the global symmetries.

Let us discuss how this $D$-flat direction is stabilized by SUSY breaking. In the $D$-flat direction, $\Sp(4)_1$ gauge symmetry is unbroken. The dynamical scale of $\Sp(4)_1$ is given as
\begin{align}
  \Lambda_{\Sp(4)} = 2^{-(N+2)/9} \left( \frac{\Lambda^{4N+11}_{2N+4}}{({\rm Pf} A \bar F \bar F)({\rm Pf} A)}  \right)^{1/9},
\end{align}
and its gaugino condensation induces the dynamical superpotential as \cite{Poppitz:1995fh, Pouliot:1995me}
\begin{align}
  W = 3 \cdot 2^{-(1+N)/3}\left( \frac{\Lambda^{4N+11}_{2N+4}}{({\rm Pf} A \bar F \bar F)({\rm Pf} A)}  \right)^{1/3}. \label{eq:Wdyn in GG model}
\end{align}
See appendix \ref{sec:Wdyn} for details.
This leads to a runaway scalar potential for $\bar F$ and $A$.
Let us introduce soft SUSY breaking masses to stabilize the vacuum:
\begin{align}
  V_{\rm soft} = m^2 |\bar F_q|^2 + m^2 |\bar F_{\bar q}|^2 + m^2 |A|^2.
\end{align}
Here we assume the universal soft SUSY breaking scalar masses.
At the global minimum of the scalar potential $V = \sum_{X=A,\bar F_q, \bar F_{\bar q}} |\partial W/\partial X|^2 + V_{\rm soft}$, $\phi$ and $c$ in eq.~(\ref{eq:vev GG model}) are given as
\begin{align}
  \phi = x_N \Lambda_{2N+4} \left( \frac{\Lambda_{2N+4}}{m}\right)^{3/(4N+8)}, \quad
  c = y_N \Lambda_{2N+4} \left( \frac{\Lambda_{2N+4}}{m}\right)^{3/(4N+8)}, \label{eq:phi and c in GG}
\end{align}
where $x_N$ and $y_N$ are ${\cal O}(1)$ coefficients determined by $N$. For its numerical value, see table~\ref{tab:xyz}.
\begin{table}
  \centering
  \begin{tabular}{|c||c|c|c|c|}
    \hline
    $N$  & $x_N$ & $y_N$ & $z_N$ & $w_N$ \\\hline\hline
    $1$  & $0.993$ & $1.23$ & $0.813$ & $0.213$\\\hline
    $3$  & $0.951$ & $1.15$ & $0.706$ & $0.233$\\\hline
    $5$  & $0.923$ & $1.10$ & $0.646$ & $0.213$\\\hline
    $10$ & $0.887$ & $1.03$ & $0.566$ & $0.172$\\\hline
    $30$ & $0.851$ & $0.941$ & $0.459$ & $0.107$\\\hline
  \end{tabular}
  \caption{$x_N$ and $y_N$ in eq.~\eqref{eq:phi and c in GG}, $z_N$ in eq.~\eqref{eq:Sp4scale in GG model}, and $w_n$ in eq.~\eqref{eq:fa in GG model} for $N = 1,~3,~5,~10,~30$.}\label{tab:xyz}
\end{table}
For $m\ll \Lambda_{2N+4}$, these VEVs are larger than $\Lambda_{2N+4}$ and it justifies our weakly coupled analysis.
Then, the dynamical scale of $\Sp(4)$ is
\begin{align}
  \Lambda_{\Sp(4)} = z_N \Lambda_{2N+4} \left( \frac{\Lambda_{2N+4}}{m}\right)^{-(2N+1)/(6N+12)}. \label{eq:Sp4scale in GG model}
\end{align}
$z_N$ is ${\cal O}(1)$ coefficient determined by $N$. For its numerical value, see table~\ref{tab:xyz}.
In addition to scalar soft masses, we can also add SUSY breaking which also violates $R$ symmetry. For example, anomaly-mediated supersymmetry breaking (AMSB) effect \cite{Randall:1998uk, Giudice:1998xp} violates SUSY and $R$ symmetry simultaneously, and it can be formulated by the Weyl compensator $\Phi = 1 + \theta^2 m_{3/2}$ \cite{Pomarol:1999ie} as
\begin{align}
  {\cal L} = \int d^2\theta \Phi \Phi^* K + \left[ \int d^2\theta \Phi^3 W + \rm{h.c.} \right].
\end{align}
At tree level, we obtain
\begin{align}
{\cal L}_{\rm AMSB} = m_{3/2} \left( -3W + \sum_i \phi_i \frac{\partial W}{\partial\phi_i}  \right) + \rm{h.c.}
\end{align}
This effect is phenomenologically important for gaugino masses and also the mass of $R$-axion. 
Note that the VEVs given in eq.~\eqref{eq:phi and c in GG} are not affected as long as $m_{3/2} \ll m$. In the limit of $m_{3/2} \gg m$, the VEVs of $\bar F_q$, $\bar F_{\bar q}$, and $A$ become close to the vacuum obtained in Ref.~\cite{Csaki:2021xhi}.

In the limit of $\alpha_{\SU(N)_2} = 0$ and $m_{3/2}=0$, the chiral Lagrangian based on $\SU(2N)_{\bar F}\times \U(1)_{\rm PQ} \times \U(1)_R / \Sp(2N)_{\bar F}$ provides a low energy effective description.
The total number of NG bosons is $2N^2-N+1$. It includes $2N^2-N-1$ NG bosons from $\SU(2N)_{\bar F}/\Sp(2N)_{\bar F}$, a PQ-axion, and an $R$-axion. Note that NG bosons in $\SU(2N)_{\bar F}/\Sp(2N)_{\bar F}$ are described as $\Yasymm$ of $\Sp(2N)_{\bar F}$. Thus, NG bosons from $\SU(2N)_{\bar F}\times \U(1)_{\rm PQ} \times \U(1)_R / \Sp(2N)_{\bar F}$ can be understood as
\begin{align}
  \Bigl[ \Yasymm \oplus \mathbf{1} \oplus \mathbf{1} \Bigr]_{\Sp(2N)_{\bar F}} = \Bigl[ \Yasymm \oplus \Yantiasymm \oplus \textbf{adj} \oplus \mathbf{1} \oplus \mathbf{1} \Bigr]_{\SU(N)_d}.
\end{align}
Here, the NG boson $\Yasymm$ has its charge $2$ under $\U(1)_d$ symmetry and $\Yantiasymm$ has the charge $-2$. The remaining NG bosons are neutral under $\U(1)_d$.
$\SU(N)_d$ gauge interaction explicitly violates the shift symmetry for those modes, and the radiative correction induces their masses \cite{Farhi:1980xs, Dobrescu:1996jp} as long as supersymmetry is broken. As a result, we obtain the masses of (pseudo) NG bosons as $m_{\rm pNGB}^2 \sim \alpha_{\SU(N)_d} m^2$.
Then, the light bosons whose masses are much smaller than $\phi$ are PQ-axion and $R$-axion. 

The mass of $R$-axion is induced by an explicit $R$ symmetry breaking. 
By using $W \sim \Lambda_{\Sp(4)}^3$, we obtain the mass of $R$-axion as
\begin{align}
  m_R^2 \sim \frac{m_{3/2} \Lambda_{\Sp(4)}^3}{\phi^2} \sim m_{3/2} m.
\end{align}
Let us discuss the domain wall number $N_{\rm DW}$ in the current model.
$\U(1)_{\rm PQ}$ transformation is given as
\begin{align}
  \bar F_q \to \bar F_q \exp\left( i(N+1)\alpha  \right), \qquad
  \bar F_{\bar q} \to \bar F_{\bar q} \exp\left( i(N+1)\alpha  \right), \qquad
  A \to A \exp\left( -iN \alpha  \right).
\end{align}
We define the anomaly coefficient of $\U(1)_{\rm PQ}$-$\SU(N)_2$-$\SU(N)_2$ as
\begin{align}
  N_{\rm anom} = \left|\sum_\Phi 2 Q_\Phi d_\Phi\right|,
\end{align}
where $Q_\Phi$ and $d_\Phi$ is the PQ charge and the Dynkin index of $\SU(N)_2$ representation of a chiral superfield $\Phi$. We take the normalization for fundamental representation to be $d = 1/2$.
Then, the anomaly coefficient is
\begin{align}
    N_{\rm anom} &= \underbrace{(N+1)(2N+4)}_{{\bar F}_q} + \underbrace{(N+1)(2N+4)}_{{\bar F}_{\bar q}} 
    = 4(N+1)(N+2).
\end{align}
$\SU(2N+4)_1$ gauge group has its center $Z_{2N+4}$ and it works as
\begin{align}
  \bar F_q \to \bar F_q \exp\left( \frac{-2\pi ik}{2N+4} \right), \quad
  \bar F_{\bar q} \to \bar F_{\bar q} \exp\left( \frac{-2\pi ik}{2N+4} \right), \quad
  A \to A \exp\left( \frac{4\pi ik}{2N+4}  \right),
\end{align}
where $k$ is an integer.
We can see that $\U(1)_{\rm PQ}$ with $\alpha = 2\pi/(N+2)$ and $Z_{2N+4}$ with $k=2$ are equivalent. Thus, to count the number of degenerated vacua $N_{\rm DW}$,
the vacua connected by a gauge transformation should be regarded as the same vacuum \cite{Lazarides:1982tw}, and we obtain
\begin{align}
  N_{\rm DW} = \frac{N_{\rm anom}}{N+2} = 4(N+1).
\end{align}

Now let us discuss the effective axion decay constant. By parametrizing ${\bar F}_q = \langle{\bar F}_q\rangle \exp\left( i(N+1)\theta \right)$, ${\bar F}_{\bar q} = \langle{\bar F}_{\bar q}\rangle \exp\left( i(N+1)\theta \right)$, and $A = \langle A\rangle \exp\left( -iN\theta \right)$, the effective Lagrangian of $\theta$ is
\begin{align}
  {\cal L} = \left(2N(N+1)^2 \phi^2 + 2c^2N^3 + 4{c'}^2N^2\right)(\partial_\mu \theta)^2 + \frac{N_{\rm anom} \theta}{32\pi^2} G_{\mu\nu}\tilde G^{\mu\nu}.
\end{align}
We define the canonically normalized axion field $a$ as
\begin{align}
  a \equiv \sqrt{4N(N+1)^2 \phi^2 + 4c^2 N^3 + 8{c'}^2 N^2} \times \theta,
\end{align}
and obtain the following effective Lagrangian:
\begin{align}
  {\cal L} &= \frac{1}{2}(\partial_\mu a)^2 + \frac{a}{f_a} \frac{1}{32\pi^2} G_{\mu\nu}\tilde G^{\mu\nu},
\end{align}
where the decay constant $f_a$ is defined as
\begin{align}
  f_a &= \frac{\sqrt{4N(N+1)^2 \phi^2 + 4c^2 N^3 + 8{c'}^2 N^2}}{N_{\rm anom}} = w_N \Lambda_{2N+4} \left( \frac{\Lambda_{2N+4}}{m}\right)^{3/(4N+8)}. \label{eq:fa in GG model}
\end{align} 
$w_N$ is coefficient determined by $N$. For its numerical value, see table~\ref{tab:xyz}.

\subsection{Axion from SUSY Bars-Yankielowicz type model}\label{sec:BY model}
\begin{table}
  \centering
\begin{tabular}{|c||c|c||c|c|c||c|}
  \hline
           & $\SU(2N-4)_1$      & $\SU(N)_2$  & $\U(1)_{\rm PQ}$ & $\U(1)_q$ & $\U(1)_R$ & $\SU(2N)_{\bar F}$ \\\hline\hline
  $\bar F_q$ & $\Yantifund$ & $\Yfund$     &  $2N-2$ & $1$ & $0$ & \multirow{2}{*}{\Yfund} \\\cline{1-6}
  $\bar F_{\bar q}$ & $\Yantifund$ & $\Yantifund$ & $2N-2$ & $-1$ & $0$ & \\\hline
  $S$      & $\Ysymm$    & $\mathbf{1}$  & $-2N$ & $0$ & $3/(N-1)$ & $\mathbf{1}$ \\\hline
  \end{tabular}
  \caption{Matter content of SUSY Bars-Yankielowicz model.}\label{tab:axion from BY model}
~\\
\begin{tabular}{|c||c|c||c|c|c||c|}
  \hline
  & $\SO(8)$ & $\SU(N)_2$ & $\U(1)_{\rm PQ}$ & $\U(1)_q$ & $\U(1)_R$ & $\SU(2N)_{\bar F}$ \\\hline\hline
$q$ & $\mathbf{8}_\mathbf{v}$ & $\Yfund$ & $-N+2$ & $1$ & $(2N-5)/(2N-2)$ & \multirow{2}{*}{$\Yantifund$} \\\cline{1-6}
$\bar q$ & $\mathbf{8}_\mathbf{v}$ & $\Yantifund$ & $-N+2$ & $-1$ & $(2N-5)/(2N-2)$ & \\\hline
$p$ & $\mathbf{8}_\mathbf{s}$ & $\mathbf{1}$ & $2N(N-2)$ & $0$ & $-(2N-5)/(N-1)$ & $\mathbf{1}$ \\\hline
$M$ & $\mathbf{1}$ & $\Ysymm$ & $2N-4$ & $2$ & $3/(N-1)$ & \multirow{4}{*}{$\Ysymm$}\\\cline{1-6}
$\bar M$ & $\mathbf{1}$ & $\overline{\Ysymm}$ & $2N-4$ & $-2$ & $3/(N-1)$ & \\\cline{1-6}
$M_a$ & $\mathbf{1}$ & $\mathbf{adj}$ & $2N-4$ & $0$ & $3/(N-1)$ & \\\cline{1-6}
$M_s$ & $\mathbf{1}$ & $\mathbf{1}$ & $2N-4$ & $0$ & $3/(N-1)$ & \\\hline
$U$ & $\mathbf{1}$ & $\mathbf{1}$ & $-4N(N-2)$ & $0$ & $6(N-2)/(N-1)$ & $\mathbf{1}$ \\\hline
\end{tabular}
\caption{Matter content of magnetic theory of SUSY Bars-Yankielowicz model. Note that $\mathbf{8}_\mathbf{v}$ and $\mathbf{8}_\mathbf{s}$ are vector and spinor representation of $\SO(8)$.}\label{tab:axion from magnetic BY model}
\end{table}

Let us discuss another example of chiral gauge theory having an axion. We introduce $\SU(2N-4)_1 \times \SU(N)_2$ gauge symmetry and chiral multiplets
$\bar F_q ~(\Yantifund, \Yfund)$,
$\bar F_{\bar q} ~(\Yantifund, \Yantifund)$,
and $S ~(\Ysymm, \mathbf{1})$.
Here we utilize the analysis presented in Ref.~\cite{Csaki:2021aqv} and assume $N\geq 9$ to have spontaneous breaking of global symmetries.
We assume that the dynamical scale of $\SU(N)_2$ gauge symmetry is much lower than that of $\SU(2N-4)_1$ gauge symmetry, i.e., $\Lambda_N \ll \Lambda_{2N-4}$. We can find $\U(1)_{\rm PQ} \times \U(1)_q \times \U(1)_{R}$ global symmetries, which are free from $\SU(2N-4)_1$ anomaly.
The matter content is summarized in table~\ref{tab:axion from BY model}.
We choose the charge assignment of $\U(1)_q$ and $\U(1)_{R}$ such that both $\U(1)_q$ and $\U(1)_{R}$ are free from $\SU(N)_2$ anomaly. We assume $W_{\rm tree} = 0$.
Note that $\SU(2N)_{\bar F}$ global symmetry arises in the limit of $\Lambda_N\to 0$ and this $\SU(2N)_{\bar F}$ global symmetry is equivalent to that of Ref.~\cite{Csaki:2021aqv}. $\SU(N)_2 \times \U(1)_q$ can be understood as a subgroup of $\SU(2N)_{\bar F}$.

The low energy degrees of freedom of this model can be described by a magnetic dual \cite{Csaki:2021aqv, Pouliot:1995sk}.
The magnetic description is given by $\SO(8)$ gauge theory.
Its matter content is summarized in table~\ref{tab:axion from magnetic BY model}.
The magnetic theory has the following superpotential:
\begin{align}
  W_{\rm mag,tree} = y_M (\tilde{\bar M} qq + \tilde M \bar q \bar q + \tilde{M_a} q \bar q + \tilde{M_s} \bar q q) + y_U \tilde U p p.
\end{align}
Note that $\tilde M$ and $\tilde U$ are normalized such that they have canonical K\"ahler potential.
We assume $y_M \sim y_U \sim 4\pi$ from naive dimensional analysis \cite{Luty:1997fk, Cohen:1997rt}.
After $q$, $\bar q$, and $p$ are integrated out, the magnetic theory becomes pure $\SO(8)$ SUSY Yang-Mills theory, and its dynamical scale $\tilde\Lambda_L$ is determined as
\begin{align}
  \tilde\Lambda^{18}_L = \frac{y_M^{2N} y_U \det \tilde {\cal M} \tilde U}{\tilde \Lambda^{2N-17}},
\end{align}
where $\tilde {\cal M}$ is defined as
\begin{align}
  \tilde {\cal M} = \left(\begin{array}{cc}
    \tilde{\bar M} & \tilde{M_a} + \tilde{M_s} \\
    (\tilde{M_a} + \tilde{M_s})^T & \tilde{M}
  \end{array}\right).
\end{align}
Thus, we obtain the dynamical superpotential as \cite{Intriligator:1995id}
\begin{align}
  W_{\rm mag,dyn} \sim \left( \frac{y_M^{2N} y_U \det \tilde {\cal M} \tilde U}{\tilde \Lambda^{2N-17}} \right)^{1/6}.
\end{align}

Here we assume SUSY breaking effect is dominated by the AMSB effect for simplicity of analysis. Let us introduce the Weyl compensator $\Phi = 1 + \theta^2 m_{3/2}$ \cite{Pomarol:1999ie} as
\begin{align}
{\cal L}_{\rm AMSB}
&= m_{3/2} \left( -3W_{\rm mag,dyn} + \sum_{\phi_i} \phi_i \frac{\partial W_{\rm mag,dyn}}{\partial\phi_i} \right) + \rm{h.c.} \nonumber\\
&\sim m_{3/2}\left( \frac{2N-17}{6}\right) W_{\rm mag,dyn} + \rm{h.c.}
\end{align}
We can find the vacuum is stabilized at
\begin{align}
  \tilde M_s \sim m_{3/2} \left( \frac{\tilde\Lambda}{m_{3/2}} \right)^{(2N-17)/(2N-11)}, \qquad
  \tilde U \sim m_{3/2} \left( \frac{\tilde\Lambda}{m_{3/2}} \right)^{(2N-17)/(2N-11)}.
\end{align}
These VEVs are smaller than $\tilde\Lambda$ and it justifies our weakly coupled analysis in the magnetic theory. Note that this vacuum is deeper than the vacuum around the origin of $\tilde {\cal M}$ and $\tilde U$ as discussed in Ref.~\cite{Csaki:2021aqv}.

The chiral Lagrangian based on $\SU(2N)_{\bar F} \times \U(1)_{\rm PQ} \times \U(1)_R / \SO(2N)_{\bar F}$ provides a low energy effective description.
The total number of NG bosons is $2N^2 + N + 1$, and it includes $2N^2 + N - 1$ NG bosons from $\SU(2N)_{\bar F} / \SO(2N)_{\bar F}$ are described as $\Ysymm$ of $\SO(2N)_{\bar F}$.
\begin{align}
  \Bigl[ \Ysymm \oplus \mathbf{1} \oplus \mathbf{1} \Bigr]_{\SO(2N)_{\bar F}} = \Bigl[ \Ysymm \oplus \Yantisymm \oplus \mathbf{adj} \oplus \mathbf{1} \oplus \mathbf{1} \Bigr]_{\SU(N)_{\bar d}}.
\end{align}
Here, the NG boson $\Ysymm$ has its charge $2$ under $\U(1)_d$ symmetry and $\Yantisymm$ has the charge $-2$. The remaining NG bosons are neutral under $\U(1)_d$.
$\SU(N)_d$ gauge interaction explicitly violates the shift symmetry for those modes, and the radiative correction induces their masses \cite{Farhi:1980xs, Dobrescu:1996jp} as long as supersymmetry is broken. As a result, we obtain the masses of (pseudo) NG bosons as $m_{\rm pNGB}^2 \sim \alpha_{\SU(N)_d} m^2$.
Then, the light bosons whose masses are much smaller than $\langle \tilde M \rangle$ are PQ-axion and $R$-axion. 
The mass of $R$-axion is estimated as
\begin{align}
  m_R^2
  \sim m_{3/2}\tilde\Lambda_L^3 \left[ m_{3/2} \left( \frac{\tilde\Lambda}{m_{3/2}} \right)^{(2N-17)/(2N-11)}\right]^{-2}
  \sim m_{3/2}^2.
\end{align}

Let us discuss the anomaly coefficient $N_{\rm anom}$ and the number of physical vacua $N_{\rm DW}$ in the current model.
In the electric theory, $\U(1)_{\rm PQ}$ transformation is given as
\begin{align}
  \bar F_q \to \bar F_q \exp\left( i(2N-2)\alpha  \right), \quad
  \bar F_{\bar q} \to \bar F_{\bar q} \exp\left( i(2N-2)\alpha  \right), \quad
  S \to S \exp\left( -2iN \alpha  \right).
\end{align}
Then, the anomaly coefficient is
\begin{align}
  N_{\rm anom}^{{\rm (ele)}} &= \underbrace{(2N-2)(2N-4)}_{{\bar F}_q} + \underbrace{(2N-2)(2N-4)}_{{\bar F}_{\bar q}} 
  = 8(N-1)(N-2).
\end{align}
$\SU(2N-4)_1$ gauge group has the center $Z_{2N-4}$ and it works as
\begin{align}
  \bar F_q \to \bar F_q \exp\left( \frac{-2\pi ik}{2N-4} \right), \qquad
  \bar F_{\bar q} \to \bar F_{\bar q} \exp\left( \frac{-2\pi ik}{2N-4} \right), \qquad
  A \to A \exp\left( \frac{4\pi ik}{2N-4}  \right),
\end{align}
where $k$ is an integer.
Therefore we can see that $\U(1)_{\rm PQ}$ with $\alpha = \pi/(N-2)$ and $Z_{2N-4}$ with $k=-2$ are equivalent. Thus, to count the number of degenerated vacua $N_{\rm vac}^{\rm (ele)}$ in the electric theory, the vacua connected by a gauge transformation should be regarded as the same vacuum \cite{Lazarides:1982tw}, and we obtain
\begin{align}
  N_{\rm DW}^{{\rm (ele)}} = \frac{N_{\rm anom}^{{\rm (ele)}}}{2(N-2)} = 4(N-1).
\end{align}
In the magnetic theory, $\U(1)_{\rm PQ}$ transformation is given as
\begin{align}
  q \to q \exp\left( -i\left(N-2 \right)\alpha \right), \quad
  p \to p \exp\left( 2iN\left(N-2 \right)\alpha \right), \quad
  M \to M \exp\left( 2i\left(N-2 \right)\alpha \right).
\end{align}
The anomaly coefficient is
\begin{align}
  N^{\rm (mag)}_{\rm anom} &= \underbrace{8(-N+2)}_{q} + \underbrace{8(-N+2)}_{\bar q} + \underbrace{(2N-4)(N+2)}_{M} + \underbrace{(2N-4)(N+2)}_{\bar M} + \underbrace{(2N-4)2N}_{M_a} \nonumber\\
  &= 8(N-1)(N-2).
\end{align}
We can find a center $Z_2 \in {\rm Spin}(8)$ which works as
\begin{align}
  q \to -q, \qquad
  p \to p.
\end{align}
Thus, $\U(1)_{\rm PQ}$ transformation with $\alpha = \pi/(N-2)$ can be identified with this $Z_2$. Then, the number of vacua in the magnetic theory is
\begin{align}
  N_{\rm DW}^{\rm (mag)} = \frac{ N_{\rm anom}^{\rm (mag)} }{2(N-2)} = 4(N-1).
\end{align}
$N_{\rm anom}^{\rm (ele)} = N_{\rm anom}^{\rm (mag)}$ can be regarded as one of anomaly matching condition between electric theory and magnetic theory.
Since $N_{\rm DW}^{\rm (ele)} = N_{\rm DW}^{\rm (mag)}$ is satisfied, the number of degenerated vacua is also consistent.

\subsection{Generalization of axion models}
So far, we have discussed two simple examples of the axion model; the SUSY Georgi-Glashow type model and the SUSY Bars-Yankielowicz type model. In both models, we identify $\bar F$ as $\Yfund + \Yantifund$ of $\SU(N)_2$ gauge theory. We can easily modify this setup by considering other representations of gauge group. For example, if we only gauge $\SU(m)$ which is a subgroup of $\SU(N)$, we can construct a Georgi-Glashow type model with $\SU(2N+4)_1 \times \SU(m)_1$ gauge symmetry and chiral multiplets $(\Yantifund,\Yfund) + (\Yantifund,\Yantifund) + (2N-2m) (\Yantifund,\mathbf{1})+ (\Yasymm,\mathbf{1})$. Another possibility is introducing larger $\SU(N)_2$ representation such as $\mathbf{adj}$.
As in discussion so far, the low energy degrees of freedom in those models can be analyzed by using chiral Lagrangian with gauge symmetry as well.

\section{GUT-motivated QCD axion model}\label{sec:GUT}
\begin{table}
  \centering
\begin{tabular}{|c||c|c||c|c||c|}
  \hline
           & $\SU(14)_1$      & $\SU(5)_2$  & $\U(1)_{\rm PQ}$ & $\U(1)_q$ & $\SU(10)_{\bar F}$\\\hline\hline
  $\bar F_q$ & $\Yantifund$ & $\Yfund$     &  $6$ & $1$ &  \multirow{2}{*}{\Yfund} \\\cline{1-5}
  $\bar F_{\bar q}$ & $\Yantifund$ & $\Yantifund$ & $6$ & $-1$ &  \\\hline
  $A$      & $\Yasymm$    & $\mathbf{1}$ & $-5$ & $0$  & $\mathbf{1}$ \\\hline\hline
  $\Phi = (\bar D \oplus L)$ & $\mathbf{1}$ & $\Yantifund$ & $0$ & $0$ & $\mathbf{1}$ \\\hline
  $\Psi = (Q \oplus \bar U \oplus \bar E)$ & $\mathbf{1}$ & $\Yasymm$ & $0$ & $0$ & $\mathbf{1}$ \\\hline
  $H$ & $\mathbf{1}$ & $\Yfund$ & $0$ & $0$ & $\mathbf{1}$ \\\hline
  $\bar H$ & $\mathbf{1}$ & $\Yantifund$ & $0$ & $0$ & $\mathbf{1}$ \\\hline
  $\Sigma$ & $\mathbf{1}$ & $\mathbf{adj}$ & $0$ & $0$ & $\mathbf{1}$ \\\hline
  \end{tabular}
  \caption{Matter content of a GUT motivated QCD axion model.}\label{tab:axion from GUT motivated model}
\end{table}
In this section, we construct a QCD axion model which is based on the SUSY Georgi-Glashow type model with $N=5$. This is a minimal setup with $\SU(5)$ GUT which is based on the discussions in the previous section.
The model has $\SU(14)_1 \times \SU(5)_2$ gauge symmetry and chiral multiplets $\bar F_q ~(\Yantifund, \Yfund)$, $\bar F_{\bar q} ~(\Yantifund, \Yantifund)$, and $A ~(\Yasymm, \mathbf{1})$. We introduce MSSM chiral multiplets as $\SU(5)_2$ charged multiplets. $\Phi~(\mathbf{1}, \Yantifund)$ behaves as $\bar D$ and $L$, $\Psi~(\mathbf{1}, \Yasymm)$ behaves as $Q$, $\bar U$, and $\bar E$. The Higgs doublets $H_u$ and $H_d$ come from $H~(\mathbf{1}, \Yfund)$ and $\bar H~(\mathbf{1}, \Yantifund)$. $\Sigma~(\mathbf{1}, \mathbf{adj})$ is introduced to break $\SU(5)_2$ gauge symmetry by its VEV\footnote{$\Sigma$ is normalized to have its K\"ahler potential as $K={\rm tr}[\Sigma^\dagger \Sigma]$.},
\begin{align}
  \langle \Sigma \rangle = \sigma\,{\rm diag}(2,2,2,-3,-3). \label{eq:VEV of Sigma}
\end{align}
The matter content and charge assignments are summarized in table~\ref{tab:axion from GUT motivated model}.
We assume that the dynamical scale of $\SU(5)_2$ gauge symmetry is much below than that of $\SU(14)_1$ gauge symmetry, i.e., $\Lambda_5 \ll \Lambda_{14}$.
The standard model gauge group $\SU(3)_c \times \SU(2)_L \times \U(1)_Y$ can be understood as a subgroup of $\SU(5)_{\rm GUT}$, which is the diagonal subgroup of $\SU(5)_1 (\subset \SU(14)_1)$ and $\SU(5)_2$.

As mentioned in section~\ref{sec:GG model}, in our model, the pseudo NG bosons~(pNGBs) appear with the PQ breaking, and are described as $\Yasymm \oplus \Yantiasymm \oplus \textbf{adj}$ of $\SU(5)_{\rm GUT}$. 
Therefore, these particles also contribute to the running of gauge couplings.
Thus, the mass of the pNGBs should be high enough to avoid the appearance of the Landau pole below the GUT scale.
On the other hand, to have a successful coupling unification, a relatively small gaugino mass is required.
Given these conditions, we assume mini-split SUSY like SUSY breaking \cite{Wells:2004di, Ibe:2006de, Hall:2011jd, Ibe:2011aa, Ibe:2012hu, Arvanitaki:2012ps, Arkani-Hamed:2012fhg} for our setup;
First, the masses of pNGBs cannot be arbitrarily large because the sfermion masses suffer from two-loop radiative correction from the pNGBs mass, and are at least one order magnitude below the pNGBs masses.
To realize the 125 GeV Higgs boson mass, the sfermion masses are at most $\sim 10^9~{\rm GeV}$. See appendix~\ref{sec:125GeV Higgs} for more details. As a result, the pNGBs masses are at most $\sim 10^{10}~{\rm GeV}$.
In the following of the analysis, we assume the sfermion masses are $\sim 10^9~{\rm GeV}$ and pNGBs masses are $\sim 10^{10}~{\rm GeV}$.
Next, we assume the gaugino masses are suppressed compared to sfermion masses. Such a mass spectrum can be naturally realized if we assume the gaugino masses are generated from AMSB effect \cite{Randall:1998uk, Giudice:1998xp}.
The masses of sfermions and pNGBs can appear from K\"ahler potential;
$K \sim Z^\dagger Z Q Q^\dagger / M_{\rm pl}^2$ as $m^2 = F_Z^2 / M_{\rm pl}^2 = {\cal O}( (10^9~{\rm GeV})^2 )$, and $K \sim Z^\dagger Z \Phi \Phi^\dagger / \Lambda^2$ as $m_{\rm pNGBs}^2 = F_Z^2 / \Lambda^2 = {\cal O}( (10^{10}~{\rm GeV})^2 )$.
The gaugino masses are suppressed by a one-loop factor as $m_{1/2} = {\cal O}(10^6~{\rm GeV})$.
We will see that this is the most favorable setup for the gauge coupling unification and Landau pole in section \ref{sec:unification}.
For this mass spectrum, the 125 GeV Higgs boson mass can be realized with $\tan\beta \sim 1$ as shown in appendix~\ref{sec:125GeV Higgs}.

The axion decay constant $f_a$ and the dynamical scale of $\Sp(4)$ $\Lambda_{\Sp(4)}$ can be written as functions of $\Lambda_{14}$ and $m$ by using eq.~(\ref{eq:fa in GG model}) and eq.~(\ref{eq:Sp4scale in GG model}) as
\begin{align}
  f_a = w_5 \Lambda_{14} \left( \frac{\Lambda_{14}}{m} \right)^{3/28}, \quad
  \Lambda_{\Sp(4)} = z_5 \Lambda_{14} \left( \frac{\Lambda_{14}}{m} \right)^{-11/42}.
  \label{eq:fa and Sp4scale in GUT model}
\end{align}
For the numerical value of $w_5$ and $z_5$, see table~\ref{tab:xyz}.

\subsection{Symmetry breaking scale and VEVs}
\label{sec:SSB and VEV}
In the current model, we have two symmetry breaking scales; the PQ breaking scale $M_{\rm PQ}$ and the GUT scale $M_{\rm GUT}$. Those two scales are determined by two VEVs of chiral multiplets. One is the VEV $\phi$ in $\bar F_q$ and $\bar F_{\bar q}$, and $A$ given in eq.~\eqref{eq:vev GG model} and this is responsible for the PQ breaking.
The other one is $\sigma$ in the VEV of $\Sigma$ given in eq.~\eqref{eq:VEV of Sigma} and this is responsible for $\SU(5)_{2/\rm GUT}$ gauge symmetry breaking. In this subsection, we outline how $M_{\rm PQ}$, $M_{\rm GUT}$, $\phi$, and $\sigma$ are related to $\Lambda_{14}$.

In our analysis, we treat $M_{\rm PQ}$ and $M_{\rm GUT}$ as energy scales at which the RG running of couplings changes.
For given $M_{\rm PQ}$, we can determine $M_{\rm GUT}$ from the following procedure.
$M_{\rm GUT}$ is determined as a scale at which two gauge couplings unify.
Suppose that there exists a solution of $M_{\rm GUT} < M_{\rm PQ}$ such that
\begin{align}
  \alpha_a(M_{\rm GUT}) = \alpha_b(M_{\rm GUT}). \label{eq:MGUT for PQ > GUT}
\end{align}
Here $\alpha_{1,2,3}$ are the gauge coupling of $\SU(3)_c \times \SU(2)_L \times \U(1)_Y$. For this definition, there are three possible choices of $(a,b)$ as $(1,2)$, $(2,3)$, and $(3,1)$. In this case, the symmetry breaking pattern from UV to IR is as follows:
\begin{align}
  &[\SU(14)_1 \times \SU(5)_2] \times \U(1)_{\rm PQ} \times \U(1)_q \times \U(1)_{R}  \nonumber\\
  \to\quad& [\Sp(4)_1 \times \SU(5)_{\rm GUT}] \times \U(1)_d \nonumber\\
  \to\quad& [\Sp(4)_1 \times \SU(3)_c \times \SU(2)_L \times \U(1)_Y] \times \U(1)_d. \label{eq:symmetry breaking pattern for PQ > GUT}
\end{align}
If there is no solution of $M_{\rm GUT} < M_{\rm PQ}$ in eq.~\eqref{eq:MGUT for PQ > GUT}, the GUT scale should be above the PQ breaking scale, i.e., $M_{\rm GUT} > M_{\rm PQ}$. In this case, the symmetry breaking pattern from UV to IR is
\begin{align}
  &[\SU(14)_1 \times \SU(5)_2] \times \U(1)_{\rm PQ} \times \U(1)_q \times \U(1)_{R}  \nonumber\\
  \to\quad& [\SU(14)_1 \times \SU(3)_2 \times \SU(2)_2 \times \U(1)_2] \times \U(1)_{\rm PQ} \times \U(1)_q \times\U(1)_R \nonumber\\
  \to\quad& [\Sp(4)_1 \times \SU(3)_c \times \SU(2)_L \times \U(1)_Y] \times \U(1)_d, \label{eq:symmetry breaking pattern for PQ < GUT}
\end{align}
and we take a tree level matching condition at $M_{\rm PQ}$ as
\begin{equation}
\begin{aligned}
  \frac{1}{\alpha_1(M_{\rm PQ})} = \frac{1}{\tilde\alpha_1(M_{\rm PQ})} + \frac{1}{\alpha_{14}(M_{\rm PQ})},\\
  \frac{1}{\alpha_2(M_{\rm PQ})} = \frac{1}{\tilde\alpha_2(M_{\rm PQ})} + \frac{1}{\alpha_{14}(M_{\rm PQ})},\\
  \frac{1}{\alpha_3(M_{\rm PQ})} = \frac{1}{\tilde\alpha_3(M_{\rm PQ})} + \frac{1}{\alpha_{14}(M_{\rm PQ})}.
  \label{eq:matching GUT>PQ}
\end{aligned}
\end{equation}
Here $\tilde\alpha_{1,2,3}$ are the gauge coupling of $\SU(3)_2 \times \SU(2)_2 \times \U(1)_2 (\subset \SU(5)_2)$ and $\alpha_{14}$ is the gauge coupling of $\SU(14)_1$.
Then, $M_{\rm GUT}$ is determined from
\begin{align}
  \tilde\alpha_a(M_{\rm GUT}) = \tilde\alpha_b(M_{\rm GUT}). \label{eq:MGUT for PQ < GUT}
\end{align}
From this procedure, we can determine $M_{\rm GUT}$ as a function of $M_{\rm PQ}$. 
Same as eq.~\eqref{eq:MGUT for PQ > GUT}, there are three possible choices of $(a,b)$ in eq.~\eqref{eq:MGUT for PQ < GUT} as $(1,2)$, $(2,3)$, and $(3,1)$.

For given $M_{\rm PQ}$ and $M_{\rm GUT}$, we can determine the VEVs $\phi$ and $\sigma$ as follows.
The gauge couplings of UV and IR theories are matched at the scale of the mass of massive gauge bosons. Thus, we determine $\phi$ as
\begin{align}
  \phi = \begin{cases}
  \left[\Tilde{g}_c^2(M_{\rm PQ}) + g_{14}^2(M_{\rm PQ})\right]^{-1/2} M_{\rm PQ} & (M_{\rm PQ} < M_{\rm GUT}) \\
  \left[(g_5^2(M_{\rm PQ}) + g_{14}^2(M_{\rm PQ})\right]^{-1/2} M_{\rm PQ} & (M_{\rm PQ} > M_{\rm GUT})
  \end{cases}\label{eq:phi from MPQ}
\end{align}
The possible choices of $c$ are $c=a$ or $c=b$, and then $\phi$ behaves as a continuous function of $M_{\rm PQ}$ at $M_{\rm PQ} = M_{\rm GUT}$.
We can also apply a similar discussion to $\sigma$ if there is a large hierarchy between $M_{\rm PQ}$ and $M_{\rm GUT}$.
For $M_{\rm PQ} \ll M_{\rm GUT}$, the VEV of $\Sigma$ given in eq.~(\ref{eq:VEV of Sigma}) breaks the $\SU(5)_{2}$ into $\SU(3)_2 \times \SU(2)_2 \times \U(1)_2$ gauge group.
Thus, the GUT scale can be understood as $M_{\rm GUT} = 5 \sqrt{2}g_{5} \sigma$. 
On the other hand, in the case of $M_{\rm PQ} \gg M_{\rm GUT}$, the GUT scale can be understood as $M_{\rm GUT} = 5 \sqrt{2}g_{\rm{GUT}} \sigma$. 
For these cases, we extract $\sigma$ as
\begin{align}
\label{eq:GUT VEV eachcase}
  \sigma = \begin{cases}
  (5\sqrt{2} g_5)^{-1} M_{\rm GUT} & (M_{\rm PQ} \ll M_{\rm GUT}) \\
  (5\sqrt{2} g_{\rm GUT})^{-1} M_{\rm GUT} & (M_{\rm PQ} \gg M_{\rm GUT})
  \end{cases}
\end{align}
So far we have discussed the cases with $M_{\rm PQ} \gg M_{\rm GUT}$ and $M_{\rm PQ} \ll M_{\rm GUT}$ to outline how the VEV of $\sigma$ is related to $M_{\rm PQ}$ and $M_{\rm GUT}$. In the next section, we discuss the case with $M_{\rm PQ} = M_{\rm GUT}$, and take the GUT scale as $M_{\rm GUT} = 5 \sqrt{2}g_{\rm{GUT}} \sigma$ as an approximation.\footnote{
Precisely speaking, this VEV $\sigma$ should behave as a continuous function of $M_{\rm PQ}$ at $M_{\rm PQ} = M_{\rm GUT}$ though eq.~\eqref{eq:GUT VEV eachcase} does not satisfy this property.
In principle, this behavior can be improved by using the fact that the running of the holomorphic coupling is one-loop exact thanks to holomorphy~\cite{Novikov:1983uc, Novikov:1985ic, Novikov:1985rd, Shifman:1986zi}.
Also, the matching condition of the holomorphic coupling is satisfied at the VEV of symmetry breaking.
Utilizing these properties, we can derive the $\sigma$ as a continuous function of $M_{\rm PQ}$.
}
Eq.~\eqref{eq:phi and c in GG} gives us the relation between $\phi$ and $\Lambda_{14}$. By using the above discussion, $M_{\rm GUT}$, $\phi$, and $\sigma$ can be expressed as functions of $M_{\rm PQ}$. Thus, by solving these relations for $\Lambda_{14}$, we can express $M_{\rm PQ}$, $M_{\rm GUT}$, $\phi$, and $\sigma$ as functions of $\Lambda_{14}$.

Let us comment on the choice of $(a,b,c)$ in eq.~\eqref{eq:MGUT for PQ > GUT}, eq.~\eqref{eq:MGUT for PQ < GUT}, and eq.~\eqref{eq:phi from MPQ}. To evaluate $M_{\rm GUT}$ and $\phi$, we have the following six choices for $(a,b,c)$: 
\begin{itemize}
 \item $(a,b,c) = (1,2,1)$
 \item $(a,b,c) = (1,2,2)$
 \item $(a,b,c) = (2,3,2)$
 \item $(a,b,c) = (2,3,3)$
 \item $(a,b,c) = (3,1,3)$
 \item $(a,b,c) = (3,1,1)$
\end{itemize}
The different choices of $(a,b)$ in eq.~\eqref{eq:MGUT for PQ > GUT} and eq.~\eqref{eq:MGUT for PQ < GUT} may lead to different $M_{\rm GUT}$ and we expect the effect of this difference should be similar to the threshold correction from the particles at the GUT scale. Also the effect of the difference between the choices of $c$ should be similar to the threshold corrections at the PQ breaking scale.
In our analysis, we are agnostic about the threshold correction, and we estimate an uncertainty of threshold correction from the difference among the choices of $(a,b,c)$.

\subsection{Numerical analysis of coupling unification}
\label{sec:unification}
In this subsection, we show a numerical analysis of the renormalization group (RG) running of the gauge couplings in the current model. As we have discussed so far, the symmetry breaking pattern from UV to IR and the RG running of the gauge couplings depend on the hierarchy between $M_{\rm PQ}$ and $M_{\rm GUT}$.
The relevant RG equations for both $M_{\rm PQ} < M_{\rm GUT}$ and $M_{\rm PQ} > M_{\rm GUT}$ are summarized in appendix \ref{sec:RGEs}.
As we will see in this section, the PQ breaking scale cannot be arbitrary low for successful GUT unification.

For our analysis, we discuss the RG running of the gauge couplings at two-loop level and the top Yukawa coupling $y_t$ at one-loop level. We neglect other Yukawa couplings.
We take the SM input parameters $g'$, $g_2$, $g_3$, and $y_t$ at $\mu = 200~{\rm GeV}$ in the $\overline{\rm MS}$ scheme from ref.~\cite{Alam:2022cdv}, and calculate the coupling constants in the $\overline{\rm DR}$ scheme. For details, see appendix \ref{sec:input parameters}.
For the MSSM parameter, we take the wino mass $M_{\tilde W} = 10^6$~GeV, the ratio of gluino and wino mass $M_{\tilde g} / M_{\tilde W} = 9$, the remaining sfermions and heavy Higgs masses to be $M_S = 2 \times 10^{9}$~GeV, the pNGBs mass $M_{\rm pNGBs} = 5 \times 10^{10}$~GeV, and $\tan\beta = 1$.
We will explain why we set this benchmark in this section.

\subsubsection{The case of $M_{\rm PQ} \geq M_{\rm GUT}$}
First, let us consider the case of $M_{\rm PQ} \geq M_{\rm GUT}$.
The symmetry breaking pattern from UV to IR is given in eq.~(\ref{eq:symmetry breaking pattern for PQ > GUT}).
At $200~{\rm GeV} < \mu < M_2$, the RG running of couplings is described by the SM beta functions given in \cite{Machacek:1983tz, Machacek:1983fi}.
Then, the SM and the SM with gauginos is matched at $\mu = M_2$ by taking care of the one-loop threshold correction~\cite{Hisano:2013cqa}.
At $M_2 < \mu < M_S$, the RG running is described by the SM + gaugino beta functions in \cite{Hisano:2013cqa}.
At $M_S < \mu < M_{\rm pNGBs}$, the RG running is described by the MSSM beta functions in eqs.~(\ref{eq:betafunction gauge MSSM}, \ref{eq:betafunction ytop MSSM}).
At $M_{\rm pNGBs} < \mu < M_{\rm GUT}$, the RG running is described by the MSSM + pNGBs beta functions in eqs.~(\ref{eq:betafunction gauge MSSM+pNGBs}, \ref{eq:betafunction ytop MSSM+pNGBs}).
At the GUT scale $M_{\rm GUT}$, $\langle\Sigma\rangle$ breaks $\SU(5)_{\rm GUT}$ gauge group into the SM gauge group as
\begin{align}
    \SU(5)_{\rm GUT} \to \SU(3)_c \times \SU(2)_L \times \U(1)_Y.
\end{align}
The tree level matching condition is given as
\begin{align}
  \frac{1}{\alpha_a(M_{\rm GUT})} = \frac{1}{\alpha_b(M_{\rm GUT})} = \frac{1}{\alpha_{\rm GUT}(M_{\rm GUT})}.
\end{align}
Here, $\alpha_{\rm GUT}$ is the gauge coupling for $\SU(5)_{\rm GUT}$. The beta functions at $M_{\rm GUT} < \mu < M_{\rm PQ}$ are given in eqs.~(\ref{eq:betafunction gauge SU(5)GUT+pNGBs}, \ref{eq:betafunction ytop SU(5)GUT+pNGBs}).
At the PQ breaking scale, the gauge symmetry is spontaneously broken as
\begin{align}
    \SU(5)_1 \times \SU(5)_2 \to \SU(5)_{\rm GUT}.
\end{align}
Then, the tree level matching condition is given as
\begin{align}
  \frac{1}{\alpha_{\rm GUT}(M_{\rm PQ})} = \frac{1}{\alpha_5(M_{\rm PQ})} + \frac{1}{\alpha_{14}(M_{\rm PQ})}.
  \label{eq:matching GUT<PQ}
\end{align}
The beta functions at $\mu > M_{\rm PQ}$ are given in eqs.~(\ref{eq:betafunction gauge SU(5)GUT+GG}, \ref{eq:betafunction ytop SU(5)GUT+GG}).

\begin{figure}[tbp]
  \centering
  \includegraphics[width=0.7\hsize]{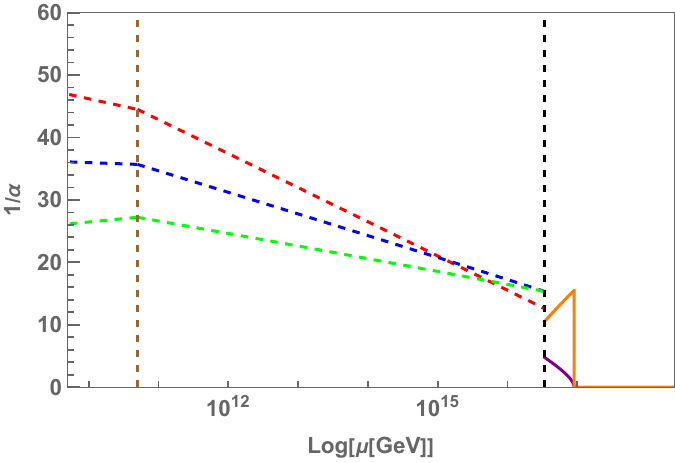}
\caption{The RG running of gauge couplings with $M_{\rm PQ} = M_{\rm GUT}$ and $(a,b) = (2,3)$. 
The red, blue, and green dashed lines show the SM gauge couplings, $\alpha_{1}$, $\alpha_2$, and $\alpha_3$, while the purple and orange solid lines denote $\alpha_{5}$ and $\alpha_{14}$, which are the gauge couplings of $\SU (5)_2$ and $\SU(14)_1$, respectively.
The vertical black and brown dashed lines depict the PQ breaking (GUT) scale and scale of the pNGBs mass, respectively.
We set the MSSM paticle mass as wino mass $M_{\tilde W} = 10^6$~GeV, the ratio of gluino and wino mass $M_{\tilde g} / M_{\tilde W} = 9$, the remaining sparticle mass $M_S = 2 \times 10^9$~GeV, the pNGBs mass $M_{\rm pNGBs} = 5 \times 10^{10}$~GeV, and $\tan\beta$ = 1. 
}
\label{fig:unification fa=MGUT}
\end{figure}

Figure \ref{fig:unification fa=MGUT} shows an example of successful GUT unification with $M_{\rm PQ} = M_{\rm GUT}$. We take the GUT scale for the case of $(a,b) = (2,3)$. The SM gauge couplings are unified successfully at the GUT (PQ breaking) scale, which is depicted by vertical black dashed line. Above that scale, the unified gauge coupling is changed to satisfy the matching conditions eq.~(\ref{eq:matching GUT<PQ}) at the PQ breaking (GUT) scale.
We will see that the case of $(a,b) = (2,3)$ can evade the current constraint on the proton decay.

In our analysis, we set the MSSM paticle mass as wino mass $M_{\tilde W} = 10^6$~GeV, the ratio of gluino and wino mass $M_{\tilde g} / M_{\tilde W} = 9$, the remaining sparticle mass $M_S = 2 \times 10^9$~GeV, the pNGBs mass $M_{\rm pNGBs} = 5 \times 10^{10}$~GeV, and $\tan\beta$ = 1.
The GUT scale and unified coupling are given by
\begin{align}
  (M_{\rm GUT},~1/\alpha_{\rm GUT})  = 
    \begin{cases}
  (1.33 \times 10^{15}~\text{GeV},~20.35) & (a,b) = (1,2) \\
  (3.41 \times 10^{16}~\text{GeV},~15.35) & (a,b) = (2,3) \\
  (5.32 \times 10^{15}~\text{GeV},~17.05) & (a,b) = (1,3) 
  \end{cases}\label{eq:Mgut aGUT}
\end{align}
In the current analysis, the GUT scale has an uncertainty of one order of magnitude as shown in the above expression.
The precise GUT scale is expected to be calculated by including threshold corrections, which are not discussed in this paper.
If the SUSY breaking scale is lower, the mass of the pNGBs is lower, then, the $1 / \alpha_{\rm GUT}$ is larger because at $M_{\rm pNGBs} < \mu < M_{\rm GUT}$, the RG running is described by the MSSM + pNGBs beta functions in eqs.~(\ref{eq:betafunction gauge MSSM+pNGBs}, \ref{eq:betafunction ytop MSSM+pNGBs}).
As a result, it is difficult to find the solution of the tree level matching condition in eq. \eqref{eq:matching GUT<PQ}. 
In addition, in the case of $M_{\rm PQ} > M_{\rm GUT}$, it becomes more difficult to satisfy eq.~\eqref{eq:matching GUT<PQ} because of the running of the $\alpha_{\rm GUT}$.
Thus, we discuss the case of $M_{\rm PQ} = M_{\rm GUT}$, and find that this benchmark is optimal for the gauge coupling unification. 

Finally, we briefly comment on the Landau pole.
As shown in figure~\ref{fig:unification fa=MGUT}, the gauge coupling hits the Landau pole below the Planck scale.
To avoid the appearance of the Landau pole, the SUSY breaking scale should be higher.
In that case, the picture of gauge coupling unification becomes worse.

\subsubsection{The case of $M_{\rm PQ} < M_{\rm GUT}$}
Next, let us consider the case of $M_{\rm PQ} < M_{\rm GUT}$.
The symmetry breaking pattern from UV to IR is given in eq.~(\ref{eq:symmetry breaking pattern for PQ < GUT}).
In this case, at the PQ breaking scale, the gauge symmetry is spontaneously broken by the VEV of $\bar F_q$ and $\bar F_{\bar q}$ as
\begin{align}
  \SU(3)_1 \times \SU(3)_2 \to \SU(3)_c, \\
  \SU(2)_1 \times \SU(2)_2 \to \SU(2)_L, \\
  \U(1)_1 \times \U(1)_2 \to \U(1)_Y.
\end{align}
Then, the tree level matching conditions at $\mu = M_{\rm PQ}$ are given as
\begin{equation}
\begin{aligned}
  \frac{1}{\alpha_1(M_{\rm PQ})} = \frac{1}{\tilde\alpha_1(M_{\rm PQ})} + \frac{1}{\alpha_{14}(M_{\rm PQ})},\\
  \frac{1}{\alpha_2(M_{\rm PQ})} = \frac{1}{\tilde\alpha_2(M_{\rm PQ})} + \frac{1}{\alpha_{14}(M_{\rm PQ})},\\
  \frac{1}{\alpha_3(M_{\rm PQ})} = \frac{1}{\tilde\alpha_3(M_{\rm PQ})} + \frac{1}{\alpha_{14}(M_{\rm PQ})}.
  \label{eq:matching GUT>PQ}
\end{aligned}
\end{equation}
Here $\tilde\alpha_{1,2,3}$ are the gauge couplings of $\SU(3)_2 \times \SU(2)_2 \times \U(1)_2 (\subset \SU(5)_2)$ and $\alpha_{14}$ is the gauge coupling of $\SU(14)_1$.
At $M_{\rm PQ} < \mu < M_{\rm GUT}$, the running of the gauge couplings is described by eqs.~(\ref{eq:betafunction gauge MSSM+GG}, \ref{eq:betafunction ytop MSSM+GG}).
As a result, in the case of $M_{\rm PQ} < M_{\rm GUT}$, we find that the Landau pole below the unification scale is inevitable, and we cannot obtain a successful GUT unification.
Thus, we do not discuss this case further.

\subsection{Proton decay}\label{sec:proton decay}
In this subsection, we discuss the proton lifetime in our model.
In the minimal SUSY SU(5) GUT, the proton decay is induced by the exchange of the color-triplet Higgs and the heavy SU(5) gauge bosons, which is described by the dimension-five and -six effective operators, respectively.
Since we assume $\mathcal{O}(10^9)$~GeV sfermion masses, the effects of the dimension-five decay operators are suppressed by the heavy sfermion masses~\cite{Hisano:2013exa, Hisano:2022qll}.
Thus, we discuss the dimension-six proton decay operators.
In our analysis, we consider the $p \to \pi^0 {e^+}$ mode, which is constrained by the experiments most severely in the case of the dimension-six proton decay operators.
In addition, we only focus on the case of $M_{\rm PQ} = M_{\rm GUT}$ because this case is optimal for the gauge coupling unification as discussed in section~\ref{sec:unification}. 

In the fermion mass basis, these relevant interactions are expressed as~\cite{Hisano:1992jj, Hisano:2012wq, Evans:2019oyw}
\begin{align}
    \mathcal{L}_{\rm int} = \frac{g_5}{\sqrt{2}} \left[- \overline{d^c_{Ri}} \gamma_\mu X_5^\mu L_i + e^{- i \varphi_i} \bar{Q'}_i \gamma_\mu X_5^\mu u^c_{Ri} + \overline{e^c_{Ri}} \gamma_\mu X_5^\mu \big(V^\dagger_{\rm CKM}\big)_{ij} Q'_j + \text{h.c.}\right],
\end{align}
where $Q'_i \equiv (u_{Li}, (V_{\rm CKM} d_L)_i )^T$, $V_{\rm CKM}$ is the Cabibbo-Kobayashi-Maskawa (CKM) matrix, $g_5$ is the SU(5) gauge coupling, $X_5^\mu$ is the SU(5) gauge bosons, and $\varphi_i$ are the GUT phase factors~\cite{Ellis:1979hy}.
By integrating out $X_5^\mu$, we can derive the dimension-six effective operators, and evolve their Wilson coefficients from $M_{\rm GUT}$ to the hadronic scale.
As a result, the partial decay width for $ p \to \pi^0 e^+$ mode is given by
\begin{align}
    \Gamma \left( p \to \pi^0 e^+ \right) = \frac{m_p}{32 \pi} \left(1 - \frac{m_\pi^2}{m_p^2}\right)^2 \left[ \left|\mathcal{A}_L \left( p \to \pi^0 e^+ \right) \right|^2 + \left|\mathcal{A}_R \left( p \to \pi^0 e^+ \right)\right|^2\right],
\end{align}
where $m_p=0.938$~GeV and $m_\pi = 0.135$~GeV are the proton and pion mass~\cite{ParticleDataGroup:2024cfk}.
The amplitudes $\mathcal{A}_L$ and $\mathcal{A}_R$ are
\begin{align}
    \mathcal{A}_L \left( p \to \pi^0 e^+ \right) &= - \frac{g_5^2 }{M_X^2}  \cdot A_1 \cdot \langle \pi^0 | (ud)_R u_L | p\rangle,\nonumber\\
    \mathcal{A}_R \left( p \to \pi^0 e^+ \right) &= - \frac{g_5^2 }{M_X^2} \cdot \left(1 + \left| V_{ud}\right|^2\right) \cdot A_2 \cdot \langle \pi^0 | (ud)_R u_L | p\rangle,
\end{align}
where $M_X$ is the mass of the SU(5) gauge bosons, and $V_{ud}$ is the $(1,1)$ element of $V_{\mathrm{CKM}}$~\cite{ParticleDataGroup:2024cfk}.
In our analysis, as seen in section~\ref{sec:SSB and VEV}, we treat $M_{\rm GUT}$ as the scale of the mass of massive gauge bosons.
The $A_1$ and $A_2$ are the renormalization factors~\cite{Abbott:1980zj, Munoz:1986kq}, and these depend on the hierarchy between $M_{\rm PQ}$ and $M_{\rm GUT}$. For details of the calculation, see appendix~\ref{sec:RGE proton decay}.
For the case of $M_{\rm PQ} = M_{\rm GUT}$, $A_1$ and $A_2$ are given by
\begin{align}
 A_1 =&
A_L \cdot 
\biggl[
\frac{\alpha_3(M_{\text{pNGBs}})}{\alpha_3(M_{\rm GUT})}
\biggr]^{- \frac{4}{15}}
\biggl[
\frac{\alpha_2(M_{\text{pNGBs}})}{\alpha_2(M_{\rm GUT})}
\biggr]^{-\frac{1}{6}}
\biggl[
\frac{\alpha_1(M_{\text{pNGBs}})}{\tilde\alpha_1(M_{\rm GUT})}
\biggr]^{-\frac{11}{438}}
\nonumber \\
&\quad\times
\biggl[
\frac{\alpha_3(M_S)}{\alpha_3(M_{\text{pNGBs}})}
\biggr]^{\frac{4}{9}}
\biggl[
\frac{\alpha_2(M_S)}{\alpha_2(M_{\text{pNGBs}})}
\biggr]^{-\frac{3}{2}}
\biggl[
\frac{\alpha_1(M_S)}{\alpha_1(M_{\text{pNGBs}})}
\biggr]^{-\frac{1}{18}}
\nonumber \\
&\quad\times
\biggl[
\frac{\alpha_3(M_Z)}{\alpha_3(M_S)}
\biggr]^{\frac{2}{7}}
\biggl[
\frac{\alpha_2(M_Z)}{\alpha_2(M_S)}
\biggr]^{\frac{27}{38}}
\biggl[
\frac{\alpha_1(M_Z)}{\alpha_1(M_S)}
\biggr]^{-\frac{11}{82}} ~, \nonumber \\
 A_2 =&
A_L \cdot
\biggl[
\frac{\alpha_3(M_{\text{pNGBs}})}{\alpha_3(M_{\rm GUT})}
\biggr]^{- \frac{4}{15}}
\biggl[
\frac{\alpha_2(M_{\text{pNGBs}})}{\alpha_2(M_{\rm GUT})}
\biggr]^{-\frac{1}{6}}
\biggl[
\frac{\alpha_1(M_{\text{pNGBs}})}{\alpha_1(M_{\rm GUT})}
\biggr]^{-\frac{23}{438}}
\nonumber \\
&\quad\times
\biggl[
\frac{\alpha_3(M_S)}{\alpha_3(M_{\rm pNGBs})}
\biggr]^{\frac{4}{9}}
\biggl[
\frac{\alpha_2(M_S)}{\alpha_2(M_{\rm pNGBs})}
\biggr]^{-\frac{3}{2}}
\biggl[
\frac{\alpha_1(M_S)}{\alpha_1(M_{\rm pNGBs})}
\biggr]^{-\frac{23}{198}}
\nonumber \\
&\quad\times
\biggl[
\frac{\alpha_3(M_Z)}{\alpha_3(M_S)}
\biggr]^{\frac{2}{7}}
\biggl[
\frac{\alpha_2(M_Z)}{\alpha_2(M_S)}
\biggr]^{\frac{27}{38}}
\biggl[
\frac{\alpha_1(M_Z)}{\alpha_1(M_S)}
\biggr]^{-\frac{23}{82}}~.
\label{eq:A MPQ<Mgut}
\end{align}
where $A_L \simeq 1.25$ is the long-distance QCD renormalization factor~\cite{Nihei:1994tx}, and $M_Z$ is $Z$ boson scale.
The $\langle\pi^0|(ud)_R{u_L}|{p}\rangle$ is the hadronic matrix element computed in lattice QCD calculations and the numerical value is~\cite{Aoki:2017puj}
\begin{align}
    \langle\pi^0|(ud)_R{u_L}|{p}\rangle = -0.131(4)(13)~\mathrm{GeV}^2.
\end{align}
In the case of $M_{\rm PQ} = M_{\rm GUT}$, we can derive the proton lifetime as
\begin{align}
  \tau_p  = 
    \begin{cases}
  2.8 \times 10^{31}~\text{years} & (a,b) = (1,2) \\
  5.4 \times 10^{36}~\text{years} & (a,b) = (2,3) \\
  4.5 \times 10^{33}~\text{years} & (a,b) = (1,3) 
  \end{cases}\label{eq:proton lifetime}
\end{align}
The current constraint on the proton lifetime of this mode is $\tau_p(p\to\pi^0{e^+})>2.4\times10^{34}$~years~\cite{Super-Kamiokande:2020wjk}, and the future sensitivity at the Hyper-Kamiokande experiment is $\tau_p(p\to\pi^0{e^+}) = 7.8 \times 10^{34}$~years~\cite{Hyper-Kamiokande:2018ofw}.
The predicted proton lifetimes in the three cases shown in eq.~\eqref{eq:proton lifetime} differ by about five orders of magnitude due to the difference in the GUT scale.
Therefore, although the predictive power is limited due to this large uncertainty, our model is potentially testable by future experiments.

\subsection{Small instanton}
Let us comment on small instanton contributions to the axion mass \cite{Holdom:1982ex, Holdom:1985vx, Dine:1986bg, Flynn:1987rs, Choi:1998ep}.
$\SU(5)_2$ gauge interaction is asymptotic non-free in the energy scale above the PQ breaking scale. Thus, we could expect that the small instanton affects the QCD axion mass. However, this contribution is extremely small as we explain below.

$\SU(5)_2$ instanton provides a 't Hooft vertex with 10 $\SU(5)_2$ gauginos, 14 $\bar{F}_q$ fermions, 14 $\bar{F}_{\bar q}$ fermions, and MSSM fermions as external legs.
Let us discuss the contraction of fermion lines of $\SU(5)_2$ gauginos, $\bar F_{q}$, and $\bar F_{\bar q}$. Total R-charge of these legs is $-18$. To have a contribution to the axion mass, we have to make contractions of fermion lines and the remaining external legs should be only PQ-charged scalar field with VEV.
Since the scalar fields of $\bar{F}_{q}$ and $\bar{F}_{\bar q}$ do not have R-charge, we need to consume 9 insertions of gaugino mass $m_\lambda$ to have only scalar field as external leg by contracting fermion lines. Thus, the axion mass squared suffers from quite strong suppression factor $(m_\lambda/\Lambda)^9 \sim 10^{-105}(m_\lambda / 1~{\rm TeV})^9 (\Lambda/M_{\rm pl})^{-9}$. Also, the additional suppression factor should come from contraction of MSSM fermions, e.g., $K \equiv \prod_q (y_q/4\pi) \sim 10^{-23}$ \cite{Agrawal:2017ksf}.
Thus, we conclude that the small instanton contribution to the axion mass is negligibly small in the current model.

\subsection{Cosmology}
The QCD axion and its superpartners play some important roles in cosmology and it highly depends on the details of the scenario. Here we briefly mention the cosmological aspects of the present model.

In the present model, the number of vacua is $N_{\rm DW } = 24$ and potentially we suffer from domain-wall problem. Furthermore, there exists unbroken $\U(1)_d$ global symmetry and some of pNGBs are charged under this $\U(1)_d$ as we have discussed in section \ref{sec:models}. Thus, the lightest pNGB with nonzero $\U(1)_d$ charge is stable. Since its masses are $\sim 10^{10}~{\rm GeV}$, this stable pNGB could overclose the universe once it is produced in the early universe.
We assume the strong CP problem is solved by the QCD axion, which implies that an explicit breaking of $\U(1)_{\rm PQ}$ symmetry is sufficiently suppressed. This immediately means an explicit breaking of $\U(1)_d$ symmetry is also strongly suppressed in our setup because $\U(1)_d$ breaking operator such as $K \supset  (A \Bar{F}_{\bar{q}}^i \Bar{F}_{\bar{q}}^j) \Phi_i \Phi_j / M_{\rm pl}^3$ breaks $\U(1)_{\rm PQ}$ simultaneously. Thus, as long as the strong CP problem is solved by the QCD axion, the good quality of $\U(1)_d$ symmetry is an inevitable consequence in our model. Therefore, the lightest pNGBs with nonzero $\U(1)_d$ charge is a stable particle.
We assume the reheating temperature is lower than $10^{10}~{\rm GeV}$ to avoid the overclosure of the universe by the pNGBs.

The QCD axion is a good candidate of the dark matter. As we have discussed, the decay constant of the axion in the current model is around GUT scale, which is higher than the canonical value of $f_a \sim 10^{12}~{\rm GeV}$ for the misalignment scenario. See, e.g., Ref.~\cite{Kawasaki:2013ae}. Thus, we need to assume that the initial misalignment angle is suppressed by some mechanism or entropy production between the onset of the axion coherent oscillation and Big Bang nucleosynthesis diluted the energy density of the axion \cite{Kawasaki:1995vt, Banks:2002sd, Kawasaki:2015pva}.

\section{Conclusions and discussions}
\label{sec:summary}
In this paper, we have discussed axion models based on supersymmetric chiral gauge theories; SUSY Georgi-Glashow type model and SUSY Bars-Yankielowicz type model. Thanks to small SUSY soft breaking, we can apply the analysis presented in Refs.~\cite{Csaki:2021xhi, Csaki:2021aqv} and we found that, in both models, spontaneous breaking of PQ symmetry is induced by non-perturbative dynamics. 
We have also discussed a GUT-motivated QCD axion model.
We have found that in order to realize the gauge coupling unification with a certain precision, the GUT scale is the same with the PQ breaking scale, and the SUSY breaking scale is ${\cal O} (10^9)~{\rm GeV} $.
Also, although the predictive power is limited due to this large uncertainty, our model is potentially testable by future experiments.

In the closing of this paper, we briefly mention non-supersymmetric limit of the current model in which all of sfermions and gauginos are decoupled. This limit is interesting because gauge invariant PQ charged scalar operator consists of at least six fermion fields and it has at least dimension 9. 
Thus, the PQ symmetry arises as an accidental symmetry and the axion quality problem can be relaxed. However, in such a regime, supersymmetry is badly broken and it is quite difficult to study in the current method. See, e.g., a discussion in section 2.4 of Ref.~\cite{Bolognesi:2021jzs}. There has been literature which discuss such a model by using the most attractive channel (MAC) analysis \cite{Raby:1979my, Dimopoulos:1980hn} and it predicts a different IR picture; 't Hooft anomaly matching condition is satisfied by not axion but massless baryon (see, e.g., \cite{Appelquist:2000qg}). To have definite conclusions in this limit, we would need a non-perturbative analysis such as lattice gauge theory simulations.

\textit{Note added:} As this paper was being completed we became aware of overlapping work in preparation from another group \cite{Gherghetta:2025kff}.

\section*{Acknowledgements}
RS thanks Pablo Qu\'ilez for useful discussions.
The work of RS is supported in part by JSPS KAKENHI Grant Numbers~23K03415, 24H02236, and 24H02244.
The work of ST is supported by JST SPRING, Grant Number JPMJSP2132.

\appendix
\section{Dynamical superpotential in SUSY Georgi-Glashow type model}\label{sec:Wdyn}
In this appendix, we discuss the dynamical superpotential in $\SU(2N)$ chiral gauge theory discussed in Ref.~\cite{Poppitz:1995fh, Pouliot:1995me}. As far as we know, the explicit coefficient of the dynamical superpotential has not given in the literature though it can be calculated in the same way as the coefficient of ADS superpotential \cite{Affleck:1983mk}.

\begin{table}
  \centering
  \begin{tabular}{|c||c|c|c|c|c|}
    \hline
    & $\SU(2N)$ & $\SU(2N-4)$ & $\U(1)_A$ & $\U(1)$ & $\U(1)_R$ \\\hline
    $\bar F$ & $\Yantifund$ & $\Yfund$ & $1$ & $2N-2$ & $1-\frac{2N}{N-2}$\\\hline
    $A$ & $\Yasymm$ & $1$ & $0$ & $-2N+4$ & $1$ \\\hline\hline
    $\Lambda^{b_{\SU(2N)}}$ & $1$ & $1$ & $2N-4$ & $2N$ & $0$ \\\hline
    ${\rm Pf} A$ & $1$ & $1$ & $0$ & $N(-2N+4)$ & $N$\\\hline
    ${\rm Pf} X$ & $1$ & $1$ & $2N-4$ & $N(2N-4)$ & $-N-6$\\\hline
  \end{tabular}
  \caption{The matter content in SUSY Georgi-Glashow model.}\label{tab:SUSY GG model}
\end{table}

We introduce $2N-4$ flavors of anti-fundamental chiral multiplets $\bar F~(\Yantifund)$ and an antisymmetric tensor chiral multiplet $A~(\Yasymm)$. Then, we can define the following gauge-invariant flavor singlet chiral fields:
\begin{align}
    {\rm Pf} A &\equiv \frac{1}{2^N N!} \epsilon^{i_1\cdots i_{2N}} A_{i_1 i_2} \cdots A_{i_{2N-1} i_{2N}}, \\
    {\rm Pf} X &\equiv \frac{1}{2^{N-2} (N-2)!} \epsilon^{a_1\cdots a_{2N-4}} X_{a_1 a_2} \cdots X_{a_{2N-5} a_{2N-4}}.
\end{align}
where $X_{ab} \equiv A_{ij} \bar F^{i}_a \bar F^{j}_b$. Note that $a$ and $b$ are flavor indices and they run from $1$ to $2N-4$, and $i$ and $j$ are gauge indices and they run from $1$ to $2N$.
These fields have global symmetries and their charges are summarized in table~\ref{tab:SUSY GG model}.
By using $\U(1) \times \U(1)_{R}$ symmetry and a spurious $\U(1)_A$ symmetry, we can write the dynamical superpotential at low energy as
\begin{align}
  W = c \left( \frac{\Lambda^{b_{\SU(2N)}}_{\SU(2N)}}{({\rm Pf}A)({\rm Pf}X)} \right)^{1/3}, \label{eq:Wdynamical}
\end{align}
where $b_{\SU(2N)}$ is the one-loop beta function coefficient given as
\begin{align}
  b_{\SU(2N)} &= 3 \times 2N - \underbrace{(2N-4) \times \frac{1}{2}}_{\bar F} - \underbrace{\frac{1}{2}(2N-2)}_{A} = 4N+3.
\end{align}
$c$ is a constant which we determine in the following of this appendix.

By assuming ${\rm Pf} A\neq 0$ and ${\rm Pf}X \neq 0$, the superpotential eq.~(\ref{eq:Wdynamical}) can be understood as a result of gaugino condensation of unbroken gauge symmetry. For simplicity of the analysis, we assume the VEV of $A$ and $\bar F$ as
\begin{align}
  \langle A \rangle = v_A (i\sigma^2 \otimes I_N), \qquad
  \langle \bar F \rangle = v_{\bar F} \left(\begin{array}{cc}
    I_{N-4} \\
    0
  \end{array}\right),
\end{align}
with $v_{\bar F} \ll v_A$. Note that $A$ and $\bar F$ are canonically normalized as $K = A_{ij} A^{* ij} + \bar F^i_a \bar F^{*a}_i$ with $A_{ij} = -A_{ji}$.
Because of the hierarchy $v_{\bar F} \ll v_A$, we can understand the spontaneous breaking of gauge symmetry as two steps. First, $\langle A \rangle$ breaks $\SU(2N)$ gauge symmetry into $\Sp(2N)$. Then, at lower energy scale, $\langle \bar F \rangle$ breaks $\Sp(2N)$ gauge symmetry into $\Sp(4)$.

Below the energy scale of $v_A$ (but still above $v_{\bar F}$), the low energy effective theory is described by $\Sp(2N)$ gauge theory. Let us discuss the matching condition between $\Lambda_{\SU(2N)}$ and $\Lambda_{\Sp(2N)}$.
$\SU(2N)$ gauge theory has $4N^2-1$ gauge bosons and $\Sp(2N)$ gauge theory has $2N^2+N$ gauge bosons. $A$ has $2N^2-N$ degrees of freedom in total. Among them, $2N^2-N-1$ degrees of freedom in $A$ are eaten by the massive gauge bosons, and the remaining one degree of freedom in $A$ parametrize the direction of $v_A$. $\bar F$ behaves as fundamental representation under $\Sp(2N)$ gauge symmetry. The massive gauge bosons form $\Yasymm$ of $\Sp(2N)$ and their mass is $m_W = 2g v_A$.
The matching condition between two holomorphic couplings is
\begin{align}
  \left( \frac{\Lambda_{\SU(2N)}}{2v_A} \right)^{b_{\SU(2N)}} = \left( \frac{\Lambda_{\Sp(2N)}}{2v_A} \right)^{b_{\Sp(2N)}}.
\end{align}
See e.g. \cite{Antoniadis:1982vr, Dine:1994su} and footnote 13 in \cite{Terning:2003th}. The coefficient $b_{\Sp(2N)}$ in one-loop beta function is given as
\begin{align}
  b_{\Sp(2N)} &= \frac{3}{2}(2N+2) - \underbrace{(2N-4) \times \frac{1}{2}}_{\bar F} = 2N+5.
\end{align}
Then, we obtain the matching condition:
\begin{align}
  \Lambda_{\Sp(2N)}^{2N+5} = \frac{ \Lambda_{\SU(2N)}^{4N+3} }{\left( 2v_A \right)^{2N-2}}. 
\end{align}

Below the energy scale of $v_{\bar F}$, the low energy effective theory is described by $\Sp(4)$ gauge theory. Let us discuss the matching condition between $\Lambda_{\Sp(2N)}$ and $\Lambda_{\Sp(4)}$.
$\Sp(2N)$ gauge theory has $2N^2+N$ gauge bosons and $\Sp(4)$ gauge theory has $10$ gauge bosons.
$\bar F$ has $4N^2-8N$ degrees of freedom in total. Among them, $2N^2 + N - 10$ degrees of freedom are eaten by the massive gauge bosons, and $2N^2-9N+9$ are NG bosons in $\SU(2N-4)/\Sp(2N-4)$, and the remaining one parameterizes in the direction of $v_{\bar F}$.
Massive gauge boson form $(2N-4) \times \Yfund + (2N^2-7N+6) \times \textbf{1}$ of $\Sp(4)$, and the mass of $\Yfund$ massive gauge boson is $gv_{\bar F}/\sqrt{2}$.
The coefficient $b_{\Sp(4)}$ in one-loop beta function is given as
\begin{align}
  b_{\Sp(4)} = \frac{3}{2} \times 6 = 9.
\end{align}
The matching condition between two holomorphic couplings is
\begin{align}
  \left( \frac{\Lambda_{\SU(2N)}}{v_{\bar F}/\sqrt{2}} \right)^{b_{\SU(2N)}} = \left( \frac{\Lambda_{\Sp(4)}}{v_{\bar F}/\sqrt{2}} \right)^{b_{\Sp(4)}}.
\end{align}
Thus, we obtain the matching condition:
\begin{align}
  \Lambda_{\Sp(4)}^{9} = \frac{ \Lambda_{\Sp(2N)}^{2N+5} }{\left( v_{\bar F}/\sqrt{2} \right)^{2N-4}}.
\end{align}
This matching condition is consistent with the condition given in \cite{Intriligator:1995ne}.
To summarize, in the limit of $v_{\bar F} \ll v_A$, we obtain
\begin{align}
  \Lambda_{\Sp(4)}^{9} = 2^{-N} \frac{ \Lambda_{\SU(2N)}^{4N+3} }{ v_{\bar F}^{2N-4} v_A^{2N-2} }.
\end{align}

Note that ${\rm Pf}A = v_A^N$, ${\rm Pf}X = v_A^{N-2} v_{\bar F}^{2N-4}$. The generic matching condition between $\Lambda_{\SU(2N)}$ and $\Lambda_{\Sp(4)}$ is
\begin{align}
  \Lambda_{\Sp(4)}^{9} = 2^{-N} \frac{ \Lambda_{\SU(2N)}^{4N+3} }{ ({\rm Pf}A)({\rm Pf}X) }.
\end{align}
For $\Sp(2N)$ pure SUSY Yang-Mills theory, its gaugino condensation induces the effective superpotential as $W = (N+1) 2^{(N-1)/(N+1)} \Lambda^3_{\Sp(2N)}$ \cite{Finnell:1995dr, Intriligator:1995ne}.
Thus, we obtain
\begin{align}
  W = 3 \cdot 2^{(1-N)/3} \left( \frac{ \Lambda_{\SU(2N)}^{4N+3} }{ ({\rm Pf}A)({\rm Pf}X) } \right)^{1/3}.
\end{align}
We can see the dynamical superpotential given in eq.~\eqref{eq:Wdyn in GG model} is equivalent to the above superpotential.

\section{The 125 GeV Higgs boson mass}
\label{sec:125GeV Higgs}

\begin{figure}[tbp]
  \centering
  \includegraphics[width=0.7\hsize]{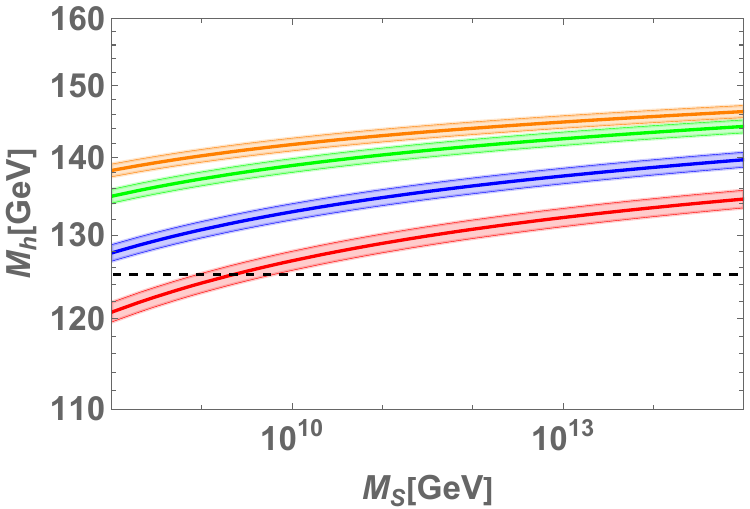}
\caption{
The Higgs boson mass $M_h$ as a function of the SUSY breaking scale $M_S$.
Red, blue, green, and yellow curves indicate the results for $\tan \beta = 1, 2, 4$, and $50$, respectively, and each color bands denote the uncertainty in the top quark mass,~$M_t = 172.4 \pm 0.7$~GeV, which we take from \cite{ParticleDataGroup:2024cfk}.
We set the wino mass as $M_{\tilde W} = 10^6$~GeV, and the ratio of gluino and wino mass as $M_{\tilde g} / M_{\tilde W} = 9$.
The black dashed line depicts the observed Higgs boson mass, 125.2 GeV~\cite{ParticleDataGroup:2024cfk}.
}
\label{fig:125GeVHiggs}
\end{figure}
In this appendix, we discuss the Higgs boson mass in our model.
As discussed in section \ref{sec:GUT}, since the mass of SUSY particles should be much higher than the electrweak scale, the radiative correction to the Higgs boson mass is significant and the leading log corrections have to be resummed.
In this appendix, we calculate the Higgs boson mass by utilizing the RGEs according to ref.~\cite{Giudice:2011cg}.

Below the mass scale of the SUSY particles, the SM gives the effective description, and the potential for the Higgs doublet $H$ is given by
\begin{align}
    V(H) = \frac{\lambda}{2} \left( H^\dagger H - v_h^2 \right)^2,
\end{align}
where $v_h$ is the VEV, and the Higgs boson mass is $M_h^2 = 2 \lambda v_h^2$.
We take a tree-level matching with supersymmetric Lagrangian at the scale of SUSY particle mass $M_S$.
Then, a matching condition for the Higgs quartic coupling $\lambda$ is
\begin{align}
\label{eq:lambda}
    \lambda (M_S) = \frac{1}{4} \left[ \frac{3}{5} g_1^2 (M_S) + g_2^2 (M_S) \right] \cos^2 2 \beta,
\end{align}
where $g_1$ and $g_2$ are $\U(1)_Y$ and $\SU(2)_L$ gauge couplings, and $\beta$ parametrizes the ratio of VEV of the Higgs fields as $\tan\beta = \langle H_u \rangle / \langle H_d \rangle$.
For other SM couplings such as $g_1$, $g_2$, and $y_t$, we fix those values of the scale of the top quark mass. The uncertainty from the top quark mass measurement $M_t = 172.4 \pm 0.7$~GeV~\cite{ParticleDataGroup:2024cfk} is taken into account following ref.~\cite{Buttazzo:2013uya}.
%
The Higgs boson mass is evaluated as $M_h^2 = 2 \lambda(M_t) v_h^2$ by using $\lambda$ at the scale of the top quark mass. This $\lambda(M_t)$ is obtained by running the coupling by the RGEs given in ref.~\cite{Giudice:2011cg}. This calculation should be consistent with the boundary condition of $\lambda$ at $\mu = M_S$ and $g_1$, $g_2$, and $y_t$ at $\mu = M_t$. For this purpose, we iteratively calculate the running of the couplings. Here we neglect threshold corrections at both the high and weak scales for simplicity.
%

Figure~\ref{fig:125GeVHiggs} shows the Higgs boson mass $M_h$ as a function of the SUSY breaking scale. 
We set the wino mass as $M_{\tilde W} = 10^6$~GeV, and the ratio of gluino and wino mass as $M_{\tilde g} / M_{\tilde W} = 9$.
In the case of $\tan \beta = 1$, we find that the 125 GeV Higgs boson mass can be realized with the SUSY breaking scale $M_S \approx 2 \times 10^9$~GeV.
In this parameter set, the gaugino masses are between the electroweak scale and the sfermion mass scale $M_S$. Thus, our prediction on $M_h$ is between the prediction of $M_h$ in Split SUSY and the high-scale SUSY scenario discussed in \cite{Giudice:2011cg} with the same $M_S$.

\section{RGEs}\label{sec:RGEs}
In this appendix, we summarize the values of the input parameters and the RGEs in the MSSM, the minimal $\SU(5)$ GUT, the MSSM with Georgi-Glashow type model, and the minimal $\SU(5)$ GUT with Georgi-Glashow type model. We show the two-loop beta functions for the gauge couplings and the one-loop beta functions for the top Yukawa coupling. We neglect other Yukawa couplings. The beta function coefficients are obtained from the generic formula given in ref.~\cite{Martin:1993zk}.

\subsection{Input parameters} \label{sec:input parameters}
Here we list the values of input parameters used for the analysis of RG running for gauge couplings and the top Yukawa coupling.
We use the values of the input parameters for gauge couplings $g'$, $g_2$, and $g_3$ and the top Yukawa coupling $y_t$ at the renormalization scale $\mu = 200~{\rm GeV}$ in the $\overline{\rm MS}$ scheme, which we take from ref.~\cite{Alam:2022cdv}.
Summarize these values in table~\ref{tab:input parameters}.

In our analysis, we use the $\overline{\rm DR}$ scheme~\cite{Siegel:1979wq}.
Then, we show the relations between the gauge couplings in the $\overline{\rm MS}$ and $\overline{\rm DR}$ scheme at one-loop level as follows~\cite{Antoniadis:1982vr, Martin:1993yx}:
\begin{align}
    g_a(\mu)_{\overline{\rm MS}} = g_a(\mu)_{\overline{\rm DR}} \left( 1 - \frac{C(G_a)}{96 \pi^2} g_a^2(\mu)_{\overline{\rm MS}}\right).
\end{align}
Here $C(G_a)$ is the quadratic Casimir invariant for the adjoint representations of group $G_a$.
On the other hand, the relation between the Yukawa couplings in the $\overline{\rm MS}$ and $\overline{\rm DR}$ scheme at one-loop level is given as~\cite{Martin:1993yx}:
\begin{align}
    (Y^{ij}_k)_{\overline{\rm MS}} = (Y^{ij}_k)_{\overline{\rm DR}} \left[1 + \sum_{a = 1}^3 \frac{g^2}{32 \pi^2} \left\{ C(r_i) + C(r_j) - 2 C(r_k) \right\}\right].
\end{align}
Here $C(r_i)$ is the quadratic Casimir invariant for the field with the subscript $i$.

\begin{table}
  \centering
\begin{tabular}{|c||c|c||c|c|c||c|}
  \hline
           & values\\\hline
  $g_3$ & $1.1525136966$\\\hline
  $g_2$ & $0.64683244428$\\  \hline
  $g'$ & $0.35885152738$\\  \hline
  $y_t$ & $0.92377763013$\\  \hline
  \end{tabular}
  \caption{The list of input parameters}\label{tab:input parameters}
\end{table}

\subsection{MSSM}\label{sec:MSSM}
\begin{table}
  \centering
  \begin{tabular}{|c||c|c|c|}
    \hline
     & ${\rm U}(1)_Y$ & $\SU(2)_L$ & $\SU(3)_c$ \\\hline\hline
    $Q_i$ & $1/6$ & $\Yfund$ & $\Yfund$ \\\hline
    $\bar U_i$ & $-2/3$ & $1$ & $\Yantifund$ \\\hline
    $\bar D_i$ & $1/3$ & $1$ & $\Yantifund$ \\\hline
    $L_i$ & $-1/2$ & $\Yfund$ & $1$ \\\hline
    $\bar E_i$ & $1$ & $1$ & $1$ \\\hline\hline
    $H_u$ & $1/2$ & $\Yfund$ & $1$ \\\hline
    $H_d$ & $-1/2$ & $\Yfund$ & $1$ \\\hline
  \end{tabular}
  \caption{The matter content in MSSM.}\label{tab:MSSM}
\end{table}
Here we discuss the MSSM whose matter content is given in table~\ref{tab:MSSM}. We introduce the top Yukawa coupling as $W = y_t Q_3 \bar U_3 H_u$.
The two-loop RGEs for the gauge couplings are given as
\begin{align}
    \frac{dg_a}{d\log\mu} &= \frac{1}{16\pi^2} b_a^{(1)} g_a^3 + \frac{g_a^3}{(16\pi^2)^2} \left[ \sum_b b_{ab}^{(2)} g_b^2 - c_a y_t^2  \right]. \label{eq:betafunction gauge MSSM}
\end{align}
$a$ runs over $1,2,3$ for $\U(1)_Y$, $\SU(2)_L$, and $\SU(3)_c$, respectively. The beta function coefficients are given as
\begin{align}
  b_a^{(1)} = \left(\begin{array}{c}
    33/5 \\ 1 \\ -3
  \end{array}\right), \qquad
  b_a^{(2)} = \left(\begin{array}{ccc}
    199/25 & 27/5 & 88/5 \\
    9/5 & 25 & 24 \\
    11/5 & 9 & 14
  \end{array}\right), \qquad
  c_a = \left(\begin{array}{c}
    26/5 \\ 6 \\ 4
  \end{array}\right). 
\end{align}
The one-loop beta function of the top Yukawa coupling is given as
\begin{align}
    \frac{dy_t}{d\log\mu} &= \frac{y_t}{16\pi^2} \left[ 6 y_t^2 - \frac{13}{15} g_1^2 - 3 g_2^2 - \frac{16}{3} g_3^2 \right]. \label{eq:betafunction ytop MSSM}
\end{align}
See also Ref.~\cite{Martin:1993zk} for details of beta functions in MSSM.

\subsection{MSSM with pNGBs}\label{sec:MSSM+pNGBs}
Here we discuss the MSSM with pNGBs.
The pNGBs are given by
\begin{align}
  \Bigl[ \Yasymm \Bigr]_{\SU(5)} &= (3,2,1/6) \oplus (\Bar{3}, 1, -2/3) \oplus (1, 1, 1),\\
  \Bigl[ \Yantiasymm \Bigr]_{\SU(5)} &= (\Bar{3},2,-1/6) \oplus (3, 1, 2/3) \oplus (1, 1, -1),\\
  \Bigl[ \textbf{adj} \Bigr]_{\SU(5)} &= (8,1,0) \oplus (1,3,0) \oplus (1,1,0) \oplus (3,2,-5/6) \oplus (\Bar{3},2,5/6).
\end{align}
We introduce the top Yukawa coupling as $W = y_t Q_3 \bar U_3 H_u$.
The two-loop RGEs for the gauge couplings are given as
\begin{align}
    \frac{dg_a}{d\log\mu} &= \frac{1}{16\pi^2} b_a^{(1)} g_a^3 + \frac{g_a^3}{(16\pi^2)^2} \left[ \sum_b b_{ab}^{(2)} g_b^2 - c_a y_t^2  \right], \label{eq:betafunction gauge MSSM+pNGBs}
\end{align}
where
\begin{align}
  b_a^{(1)} = \left(\begin{array}{c}
    73/5 \\ 9 \\ 5
  \end{array}\right), \qquad
  b_a^{(2)} = \left(\begin{array}{ccc}
    1567/75 & 21 & 808/15 \\
    7 & 91 & 56 \\
    101/15 & 21 & 374/3
  \end{array}\right), \qquad
  c_a = \left(\begin{array}{c}
    26/5 \\ 6 \\ 4
  \end{array}\right). 
\end{align}
The one-loop beta function of the top Yukawa coupling is given as
\begin{align}
    \frac{dy_t}{d\log\mu} &= \frac{y_t}{16\pi^2} \left[ 6 y_t^2 - \frac{13}{15} g_1^2 - 3 g_2^2 - \frac{16}{3} g_3^2 \right]. \label{eq:betafunction ytop MSSM+pNGBs}
\end{align}

\subsection{Minimal $\SU(5)$ GUT}\label{sec:minimal SU(5) GUT}
\begin{table}
  \centering
  \begin{tabular}{|c||c|}
    \hline
     & $\SU(5)_{\rm GUT}$ \\\hline\hline
    $\Phi_i$ & $\Yantifund$ \\\hline
    $\Psi_i$ & $\Yasymm$ \\\hline\hline
    $H$ & $\Yfund$ \\\hline
    $\bar H$ & $\Yantifund$ \\\hline
    $\Sigma$ & $\mathbf{adj}$  \\\hline
  \end{tabular}
  \caption{The matter content in the minimal $\SU(5)$ GUT.}\label{tab:minimalGUT}
\end{table}
Here we discuss the minimal SU(5) GUT whose matter content is given in table~\ref{tab:minimalGUT}.
We introduce the top Yukawa coupling as $W = y_t \Psi_3 \Psi_3 H$. We do not include $\Sigma^3$ and $H\Sigma\bar H$ coupling in the superpotential.
The two-loop beta function for the gauge coupling is given as
\begin{align}
  \frac{dg_5}{d\log\mu} = \frac{1}{16\pi^2} b^{(1)} g_5^3 + \frac{g_5^3}{(16\pi^2)^2} \left[ b^{(2)} g_5^2 - c y_t^2  \right], \label{eq:betafunction gauge SU(5)GUT}
\end{align}
where
\begin{align}
  b^{(1)} = -3, \qquad
  b^{(2)} = 794/5, \qquad
  c = 12.
\end{align}
The one-loop beta function of the top Yukawa coupling is given as \cite{Hisano:1992jj, Wright:1994qb}
\begin{align}
  \frac{dy_t}{d\log\mu} &= \frac{y_t}{16\pi^2} \left[ 9 y_t^2 - \frac{96}{5} g_5^2 \right]. \label{eq:betafunction ytop SU(5)GUT}
\end{align}

\subsection{Minimal $\SU(5)$ GUT with pNGBs}\label{sec:minimal SU(5) GUT+pNGBs}

Here we discuss the minimal SU(5) GUT with pNGBs.
The pNGBs are $\Yasymm \oplus \Yantiasymm \oplus \mathbf{Adj}$ under $\SU(5)$.
We introduce the top Yukawa coupling as $W = y_t \Psi_3 \Psi_3 H$. We do not include $\Sigma^3$ and $H\Sigma\bar H$ coupling in the superpotential.
The two-loop beta function for the gauge coupling is given as
\begin{align}
  \frac{dg_5}{d\log\mu} = \frac{1}{16\pi^2} b^{(1)} g_5^3 + \frac{g_5^3}{(16\pi^2)^2} \left[ b^{(2)} g_5^2 - c y_t^2  \right], \label{eq:betafunction gauge SU(5)GUT+pNGBs}
\end{align}
where
\begin{align}
  b^{(1)} = 5, \qquad
  b^{(2)} = 382, \qquad
  c = 12.
\end{align}
The one-loop beta function of the top Yukawa coupling is given as \cite{Hisano:1992jj, Wright:1994qb}
\begin{align}
  \frac{dy_t}{d\log\mu} &= \frac{y_t}{16\pi^2} \left[ 9 y_t^2 - \frac{96}{5} g_5^2 \right]. \label{eq:betafunction ytop SU(5)GUT+pNGBs}
\end{align}

\subsection{MSSM with Georgi-Glashow type model}\label{sec:MSSM+GG}
\begin{table}
  \centering
  \begin{tabular}{|c||c|c|c|c|}
    \hline
     & ${\rm U}(1)_Y$ & $\SU(2)_L$ & $\SU(3)_c$ & $\SU(14)_1$\\\hline\hline
    $Q_i$ & $1/6$ & $\Yfund$ & $\Yfund$ & $1$\\\hline
    $\bar U_i$ & $-2/3$ & $1$ & $\Yantifund$  & $1$\\\hline
    $\bar D_i$ & $1/3$ & $1$ & $\Yantifund$ & $1$ \\\hline
    $L_i$ & $-1/2$ & $\Yfund$ & $1$ & $1$ \\\hline
    $\bar E_i$ & $1$ & $1$ & $1$ & $1$ \\\hline\hline
    $H_u$ & $1/2$ & $\Yfund$ & $1$ & $1$ \\\hline
    $H_d$ & $-1/2$ & $\Yfund$ & $1$ & $1$ \\\hline
    $D_{14}$ & $-1/3$ & $1$ & $\Yfund$ & $\Yantifund$ \\\hline
    $\bar D_{14}$ & $1/3$ & $1$ & $\Yantifund$ & $\Yantifund$ \\\hline
    $L_{14}$ & $-1/2$ & $\Yfund$ & $1$ & $\Yantifund$ \\\hline
    $\bar L_{14}$ & $1/2$ & $\Yfund$ & $1$ & $\Yantifund$ \\\hline
    $A$ & $0$ & $1$ & $1$ & $\Yasymm$ \\\hline
  \end{tabular}
  \caption{The matter content in MSSM with Georgi-Glashow type model.}\label{tab:MSSM+GG}
\end{table}
Here we discuss the MSSM with Georgi-Glashow type model whose matter content is given in table~\ref{tab:MSSM+GG}.
We introduce the top Yukawa coupling as $W = y_t Q_3 \bar U_3 H_u$.
The two-loop beta functions for the gauge couplings are given as
\begin{align}
  \frac{dg_a}{d\log\mu} = \frac{1}{16\pi^2} b_a^{(1)} g_a^3 + \frac{g_a^3}{(16\pi^2)^2} \left[ \sum_b b_{ab}^{(2)} g_b^2 - c_a y_t^2  \right], \label{eq:betafunction gauge MSSM+GG}
\end{align}
where
\begin{align}
  b_a^{(1)} = \left(\begin{array}{c}
    103/5 \\ 15 \\ 11 \\ -31
  \end{array}\right), \qquad
  b_a^{(2)} = \left(\begin{array}{cccc}
    1087/75 & 153/5 & 712/15 & 390 \\
    51/5 & 123 & 24 & 390 \\
    89/15 & 9 & 518/3 & 390 \\
    2 & 6 & 16 & -2941/7 \\
  \end{array}\right), \qquad
  c_a = \left(\begin{array}{c}
    26/5 \\ 6 \\ 4 \\ 0
  \end{array}\right).
\end{align}
The one-loop beta function of the top Yukawa coupling is given as
\begin{align}
  \frac{dy_t}{d\log\mu} &= \frac{y_t}{16\pi^2} \left[ 6 y_t^2 - \frac{13}{15} g_1^2 - 3 g_2^2 - \frac{16}{3} g_3^2 \right]. \label{eq:betafunction ytop MSSM+GG}
\end{align}

\subsection{Minimal $\SU(5)$ GUT with Georgi-Glashow type model}\label{sec:minimal SU(5) GUT+GG}
\begin{table}
  \centering
  \begin{tabular}{|c||c|c|}
    \hline
     & $\SU(5)_2$ & $\SU(14)_1$ \\\hline\hline
    $\Phi_i$ & $\Yantifund$ & $1$\\\hline
    $\Psi_i$ & $\Yasymm$ & $1$ \\\hline\hline
    $H$ & $\Yfund$ & $1$ \\\hline
    $\bar H$ & $\Yantifund$ & $1$ \\\hline
    $\Sigma$ & $\mathbf{adj}$ & $1$ \\\hline
    $\bar F_q$ & $\Yfund$ & $\Yantifund$ \\\hline
    $\bar F_{\bar q}$ & $\Yantifund$ & $\Yantifund$ \\\hline
    $A$ & $1$ & $\Yasymm$ \\\hline
  \end{tabular}
  \caption{The matter content in the minimal $\SU(5)$ GUT with Georgi-Glashow type model.}\label{tab:minimalGUT+GG}
\end{table}
Here we discuss the minimal SU(5) GUT with Georgi-Glashow type model whose matter content is given in table~\ref{tab:minimalGUT+GG}. We introduce the top Yukawa coupling as $W = y_t \Psi_3 \Psi_3 H$. We do not include $\Sigma^3$ and $H\Sigma\bar H$ coupling in the superpotential.
The two-loop beta functions for the gauge couplings are given as
\begin{align}
  \frac{dg_a}{d\log\mu} = \frac{1}{16\pi^2} b_a^{(1)} g_a^3 + \frac{g_a^3}{(16\pi^2)^2} \left[ \sum_b b_{ab}^{(2)} g_b^2 - c_a y_t^2  \right], \label{eq:betafunction gauge SU(5)GUT+GG}
\end{align}
where
\begin{align}
  b_a^{(1)} = \left(\begin{array}{c}
    11 \\ -31
  \end{array}\right), \qquad
  b_a^{(2)} = \left(\begin{array}{cc}
    2166/5 & 390 \\
    48 & -2941/7
  \end{array}\right), \qquad
  c_a = \left(\begin{array}{c}
    12 \\ 0
  \end{array}\right).
\end{align}
The one-loop beta function of the top Yukawa coupling is given as
\begin{align}
  \frac{dy_t}{d\log\mu} &= \frac{y_t}{16\pi^2} \left[ 9 y_t^2 - \frac{96}{5} g_5^2 \right]. \label{eq:betafunction ytop SU(5)GUT+GG}
\end{align}

\section{RGEs for proton decay} \label{sec:RGE proton decay}
Here we briefly summarize the RGEs for proton decay operators. We only discuss the following dimension-six operators which are induced by the exchange of heavy gauge bosons:
\begin{align}
    {\cal L}_{\rm eff}
    = C^{ijkl}_{6(1)} \epsilon_{abc} (u^a_{Ri} d^b_{Rj})( q^c_{Lk} L_{Ll})
    + C^{ijkl}_{6(2)} \epsilon_{abc} (q^a_{Li} q^b_{Lj})( u^c_{Rk} e_{Rl}) + \rm{h.c.}
\end{align}
The RGEs for Wilson coefficients in the SM are given as \cite{Abbott:1980zj}
\begin{align}
  \frac{d}{d\log\mu} C^{ijkl}_{6(1)} &= \left( -\frac{11}{10}\frac{\alpha_1}{4\pi} - \frac{9}{2}\frac{\alpha_2}{4\pi} - 4 \frac{\alpha_3}{4\pi} \right) C^{ijkl}_{6(1)}, \\
  \frac{d}{d\log\mu} C^{ijkl}_{6(2)} &= \left( -\frac{23}{10}\frac{\alpha_1}{4\pi} - \frac{9}{2}\frac{\alpha_2}{4\pi} - 4 \frac{\alpha_3}{4\pi} \right) C^{ijkl}_{6(2)}.
\end{align}
Similarly, the RGEs in SUSY models (including MSSM) are \cite{Munoz:1986kq}
\begin{align}
  \frac{d}{d\log\mu} C^{ijkl}_{6(1)} &= \left( -\frac{11}{15}\frac{\alpha_1}{4\pi} - 3 \frac{\alpha_2}{4\pi} - \frac{8}{3} \frac{\alpha_3}{4\pi} \right) C^{ijkl}_{6(1)}, \\
  \frac{d}{d\log\mu} C^{ijkl}_{6(2)} &= \left( -\frac{23}{15}\frac{\alpha_1}{4\pi} - 3 \frac{\alpha_2}{4\pi} - \frac{8}{3} \frac{\alpha_3}{4\pi} \right) C^{ijkl}_{6(2)}.
\end{align}
Let us write the one-loop RGE for the gauge coupling as
\begin{align}
  \frac{d g_a}{d\log\mu} = \frac{1}{16\pi^2} b_a^{(1)} g_a^3,
\end{align}
where $a$ runs over $1,2,3$ for $\U(1)_Y$, $\SU(2)_L$, and $\SU(3)_c$, respectively.
Then, the Wilson coefficients at the scale $\mu$ and $\mu_0$ in the SM are related as
\begin{align}
  C_{6(1)}^{ijkl}(\mu) &=
  \left( \frac{\alpha_1(\mu)}{\alpha_1(\mu_0)}\right)^{-11/20b_1^{(1)}}
  \left( \frac{\alpha_2(\mu)}{\alpha_2(\mu_0)}\right)^{-9/4b_2^{(2)}}
  \left( \frac{\alpha_3(\mu)}{\alpha_3(\mu_0)}\right)^{-2/b_3^{(2)}} C_{6(1)}^{ijkl}(\mu_0), \\
  C_{6(2)}^{ijkl}(\mu) &=
  \left( \frac{\alpha_1(\mu)}{\alpha_1(\mu_0)}\right)^{-23/20b_1^{(1)}}
  \left( \frac{\alpha_2(\mu)}{\alpha_2(\mu_0)}\right)^{-9/4b_2^{(2)}}
  \left( \frac{\alpha_3(\mu)}{\alpha_3(\mu_0)}\right)^{-2/b_3^{(2)}} C_{6(2)}^{ijkl}(\mu_0).
\end{align}
Also, the Wilson coefficients at the scale $\mu$ and $\mu_0$ in SUSY models are related as
\begin{align}
  C_{6(1)}^{ijkl}(\mu) &=
  \left( \frac{\alpha_1(\mu)}{\alpha_1(\mu_0)}\right)^{-11/30b_1^{(1)}}
  \left( \frac{\alpha_2(\mu)}{\alpha_2(\mu_0)}\right)^{-3/2b_2^{(2)}}
  \left( \frac{\alpha_3(\mu)}{\alpha_3(\mu_0)}\right)^{-4/3b_3^{(2)}} C_{6(1)}^{ijkl}(\mu_0), \\
  C_{6(2)}^{ijkl}(\mu) &=
  \left( \frac{\alpha_1(\mu)}{\alpha_1(\mu_0)}\right)^{-23/30b_1^{(1)}}
  \left( \frac{\alpha_2(\mu)}{\alpha_2(\mu_0)}\right)^{-3/2b_2^{(2)}}
  \left( \frac{\alpha_3(\mu)}{\alpha_3(\mu_0)}\right)^{-4/3b_3^{(2)}} C_{6(2)}^{ijkl}(\mu_0).
\end{align}

\bibliography{ref}
\bibliographystyle{JHEP}
\end{document}